\documentclass[11pt,a4paper]{article}
\pdfoutput=1
\usepackage{jheppub}

\usepackage{amsmath}
\usepackage{color}
\usepackage{graphicx}
\usepackage{url}
\usepackage{epsfig}
\usepackage[T1]{fontenc}
\usepackage[latin9]{inputenc}
\usepackage{multirow}

\def\sp{S}
\def\s0{\sigma_0}
\def\alphaEM{\alpha_{\textrm{em}}}
\def\beq{\begin{equation}}
\def\eeq{\end{equation}}
\def\bear{\begin{eqnarray}}
\def\enar{\end{eqnarray}}
\def\be{\begin{equation}}
\def\ee{\end{equation}}

\def\qperp{q_\perp}

\newcommand{\as}{\bar{\alpha}_s}
\newcommand{\half}{\textstyle{\frac{1}{2}}}

\title{Forward Drell-Yan and backward jet production as a probe of the BFKL dynamics} 

\author[a]{Krzysztof Golec-Biernat}
\author[b]{Leszek Motyka}
\author[a,c]{Tomasz Stebel}

\affiliation[a]{Institute of Nuclear Physics PAN, Radzikowskiego 152, 31-342 Krak\'ow, Poland}
\affiliation[b]{Institute of Physics, Jagiellonian University, S.\L{}ojasiewicza 11, 30-348 Krak\'ow, Poland} 
\affiliation[c]{Physics Department,Brookhaven National Laboratory,Upton, NY 11973, USA}

\emailAdd{golec@ifj.edu.pl}
\emailAdd{leszekm@th.if.uj.edu.pl}
\emailAdd{tomasz.stebel@ifj.edu.pl}
\abstract{We propose a new process which probes the BFKL dynamics in the high energy proton-proton scattering, namely the forward Drell-Yan (DY) production accompanied by a backward jet, separated from the DY lepton pair by a large rapidity interval. The proposed process probes higher rapidity differences and smaller transverse momenta than in the Mueller-Navelet jet production. It also offers a possibility of measuring new observables like lepton angular distribution coefficients in the DY lepton pair plus jet production. }

\begin{document}
\maketitle

\section{Introduction}

After almost a decade since the launch, the Large Hadron Collider (LHC) operates at energy $\sqrt{S}=13$ TeV, close to the nominal $\sqrt{S} = 14$~TeV, and the total integrated luminosities are large enough to perform precision studies of physics at electroweak scales. Currently, precision physics offers one of the most promising paths towards potential discoveries of physics beyond the Standard Model. Theoretical precision for the observables at the LHC requires good understanding of strong interactions, that govern the structure of beams, drive or introduce sizable corrections to most interesting reactions. In particular, accurate description of high energy hadronic collisions crucially depends on good understanding of color radiation and the resulting final states. In the high energy regime, the QCD radiation is intense, and the theoretical treatment requires calculational schemes that go beyond fixed order QCD calculations. In this regime, an all order resummation of the perturbative QCD corrections enhanced by powers of energy logarithms, $(\log{\sqrt{S}})^n$, is necessary that leads to the celebrated BFKL formalism \cite{Lipatov:1976zz, Kuraev:1976ge, Kuraev:1977fs, Balitsky:1978ic, Lipatov:1996ts}. This formalism is complementary to the collinear resummations scheme and is used to improve predictions for cross sections and final states in hadronic collision at high energies. 
Hence it is necessary to provide predictions for new observables that carry significant BFKL effects.

In this paper, we propose a new probe of the BFKL dynamics given by a forward Drell-Yan (DY) pair production in association with a backward jet. We closely follow the approach and methods developed for forward-backward jet hadroproduction \cite{Mueller:1986ey}, described below. 

A classical probe of QCD radiation in the BFKL approach applied to hadronic collisions was proposed by Mueller and Navelet (MN) \cite{Mueller:1986ey} to study hadroproduction of two jets with similar transverse momenta but separated by a large rapidity interval $\Delta Y$ which are produced from a collision of two partons with moderate hadron momentum fractions.
For such a configuration, the phase space for QCD radiation is large and so are the emerging logarithms of energy. The first analysis of the dijet production data from the Tevatron \cite{DelDuca:1993mn} showed that the exponential enhancement with $\Delta Y$ in dijet production at fixed parton momentum fractions, as originally suggested by Mueller and Navelet, is highly suppressed by the parton distribution functions at Tevatron energies. Thus, it was proposed in \cite{DelDuca:1993mn,Stirling:1994he,DelDuca:1994ng} to use the angular decorrelation in transverse momentum and azimuthal angle in the transverse plane of the MN jets as a new probe of the BFKL dynamics. Both observables became a subject of intense experimental studies at Tevatron \cite{Abachi:1996et,Abbott:1999ai} and the LHC \cite{Aad:2011jz, Chatrchyan:2012pb, Aad:2014pua, Khachatryan:2016udy}. From the theoretical side, a substantial theoretical progress has been made since the appearance of the initial papers to include the next-to-leading order (NLO) corrections to the jet impact factors and the next-to-leading-logarithmic (NLL) corrections to the BFKL evolution kernel to make a successful comparison with data. Below, we briefly describe this progress.

The first evidence of significant NLL effects in MN jets came from confronting the $2 \to 3$ parton fixed order calculations with the first iteration of the LL BFKL kernel \cite{DelDuca:1994ng}. Already first approaches to include leading higher order corrections to the BFKL evolutions showed that they substantially modify the LL BFKL predictions \cite{Kwiecinski:2001nh,Vera:2006un,Marquet:2007xx}. The key steps towards obtaining full NLO/NLL BFKL predictions for the MN jet observables were made by the computation of the NLL BFKL kernel \cite{Fadin:1995xg,Fadin:1996nw,Fadin:1997zv,Fadin:1997hr,Fadin:1998py}, and the computation of the quark and gluon impact factors at NLO \cite{Bartels:2001ge,Bartels:2002yj}. The first results for the MN jets with the NLL BFKL kernel, but using the LO impact factors, was presented in \cite{Vera:2007kn} and the full NLO/NLL predictions were given in \cite{Colferai:2010wu, Caporale:2011cc}. It was shown in \cite{Ducloue:2013wmi, Ducloue:2013bva, Caporale:2014gpa} that the NLO/NLL BFKL results describe well the MN jet data collected at the LHC. However, to achieve good agreement it was required to fix the process scale in the BLM procedure \cite{Brodsky:1982gc} with a surprisingly large hard scale. An interesting alternative to this procedure was proposed in \cite{Caporale:2013uva} where an all order collinear improvement was applied to the NLL BFKL kernel with a natural process scale. Finally, it was shown in \cite{Celiberto:2015yba} that the BFKL effects are clearly distinguishable from the DGLAP effects. 

The theoretical effort described above and some remaining puzzles clearly indicate the need to test the scheme with other processes. In Ref.\ \cite{Andersen:2001ja} BFKL effects were analyzed in the $W$ boson production in association with one and two jets.
Recently proposals were made to combine the backward MN jet with a forward probe being a heavy quarkonium \cite{Boussarie:2017oae}, the Higgs boson \cite{Xiao:2018esv} or the charged light hadron \cite{Bolognino:2018oth}. In this paper we propose to replace one of the MN jets by a forward Drell-Yan pair. At the partonic level, it amounts to replacing the $qg^* \to q$ impact factor by the $q g^* \to q \gamma^* \to q l^+l^-$ impact factor. There are several advantages to use the forward Drell-Yan pair as one of the probes.
(i) The experimental precision of DY measurements is usually very high. 
(ii) The forward production of the DY pair with a backward jet depends on several kinematical variables which may be scanned: mass of the lepton pair $M$, its transverse momentum $q_\perp$ and rapidity $y_{\gamma}$, and the virtual boson -- jet separation in rapidity $Y_{\gamma J}$. 
(iii) The lepton angular distributions depend on three independent coefficients related to 
the DY structure functions \cite{Lam:1978pu,Lam:1980uc,Motyka:2014lya,Motyka:2016lta,Brzeminski:2016lwh} in which
some theoretical uncertainties are expected to cancel out. 
(iv) The Lam-Tung combination of the DY structure functions \cite{Lam:1978pu,Lam:1980uc} is particularly sensitive to the transverse momentum of the exchanged $t$-channel parton \cite{Motyka:2014lya,Motyka:2016lta,Brzeminski:2016lwh}, hence to the effects of the QCD radiation in the exchange. 
Thus, given the richness of the interesting observables and their sensitivity to BFKL effects, the forward DY pair$+$backward jet production offers an excellent testing ground for theory.

In calculations of the BFKL scattering amplitudes one applies the high-energy factorization framework. 
Up to now, the forward Drell-Yan impact factors for all virtual photon polarizations are known only at the leading order \cite{Brodsky:1996nj,Kopeliovich:2000fb,Kopeliovich:2001hf,Gelis:2002fw,Motyka:2014lya,Schafer:2016qmk}, and the analogous impact factors for forward lepton hadroproduction through the $W$ boson were also calculated at the LO \cite{Andersen:2001ja}. These impact factors, combined with the LL \cite{Brzeminski:2016lwh} or NLL \cite{Celiberto:2018muu} BFKL evolution, lead to successful description of the inclusive Drell-Yan cross section at the LHC within the BFKL framework.
Since the NLO Drell-Yan impact factors are not available yet, the full NLO/NLL BFKL calculation cannot be done also for the DY\,$+$\,jet process. So we choose an approach closely following the one applied in \cite{Kwiecinski:2001nh} in which the LO impact factors are combined with the LL BFKL kernel with all order collinear improvements \cite{Andersson:1995jt, Kwiecinski:1996td, Kwiecinski:1997ee, Salam:1998tj,Ciafaloni:1999yw, Salam:1999cn}. We apply the implementation of the collinear improvements called the consistency condition, defined in \cite{Kwiecinski:1996td}. Although this simplified approach does not enjoy the theoretical sophistication of the full NLO/NLL BFKL calculations, it is expected to encompass the generic properties of the QCD radiation at high energies. In particular, it follows from \cite{Kwiecinski:1996td} and \cite{Fadin:1995xg,Fadin:1996nw,Fadin:1997zv,Fadin:1997hr,Fadin:1998py},
that the collinear improvements to the BFKL kernel, constrained to the NLL accuracy exhausts up to $70\%$ of the exact NLL BFKL corrections \cite{Kwiecinski:1998sa}. Therefore, we expect to obtain the correct indications of general phenomenological properties of the studied observables. The results obtained in this paper clearly show the significance of the BFKL effects in associated Drell-Yan and jet hadroproduction, and allow to propose this process as a sensitive probe of the BFKL dynamics.

The paper is organized as follows. In Section~\ref{sec:2} we introduce kinematic variables for the DY\,$+$\,jet process, while in Section~\ref{sec:3} we present basic formulas for the DY\,$+$\,jet cross section. In particular, we present the BFKL kernel and lepton angular distribution coefficients as well as the MN jet cross section as a handy reference. In Section~\ref{sec:6} we discuss our numerical results for the DY\,$+$\,jet process, namely the dependence on the azimuthal angle $\phi_{\gamma J}$ between the DY photon and the backward jet, which shows the angular decorrelation elaborated later in terms of the mean cosine values. We also present in this section the results on the angular coefficients of the DY leptons, which provide additional information
to be measured in the experiments. Finally, we present summary and outlook.

\section{Kinematic variables}
\label{sec:2}

\begin{figure}[t]
\begin{center}
\includegraphics[width=.37\textwidth]{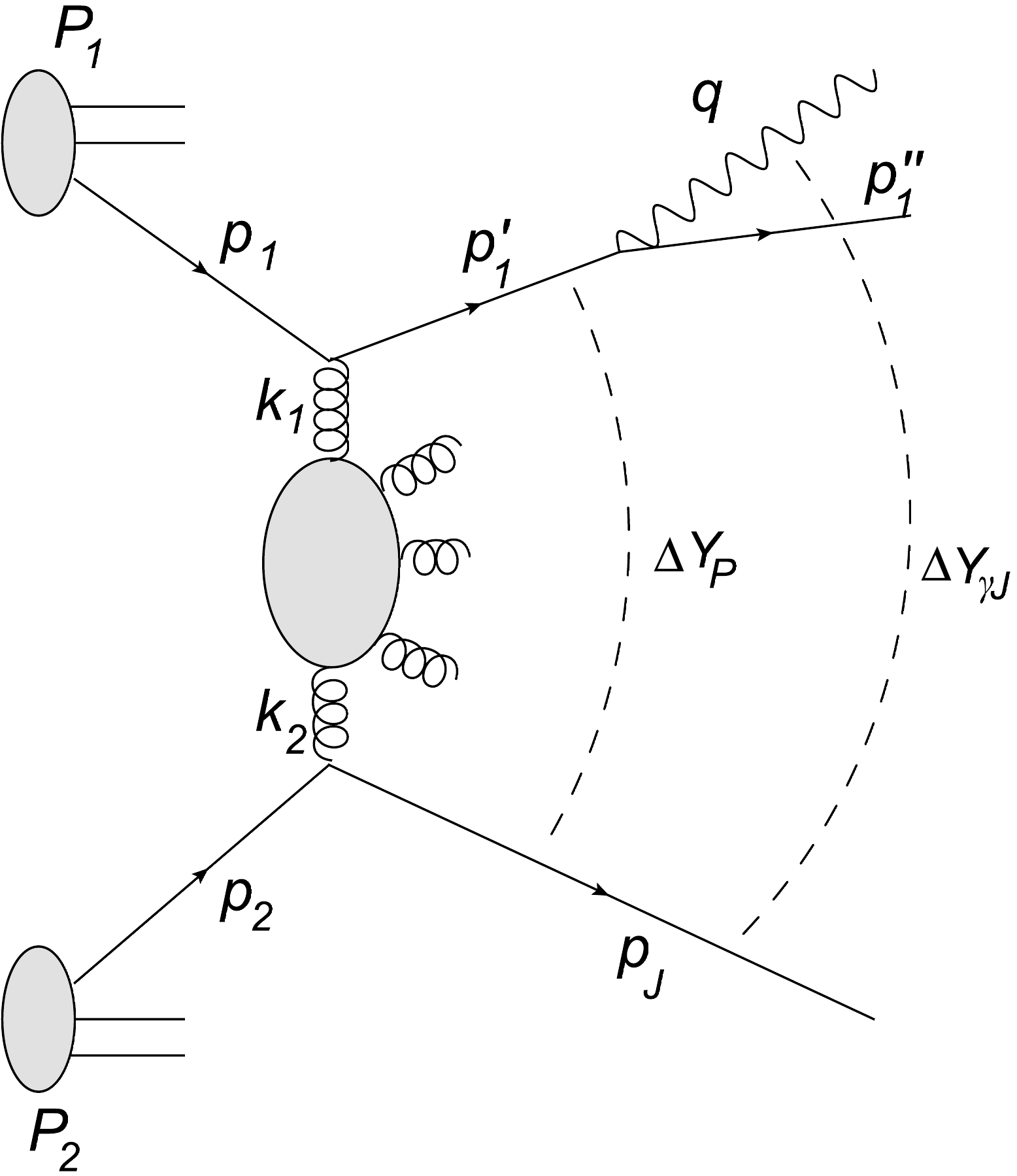}
\end{center}
\caption{One of the two diagrams for the forward Drell-Yan\,$+$\,backward jet production with the indicated kinematic variables. Only photon $q$ and jet $p_J$ momenta are measured. Parton $p_2$ might be either quark or gluon. 
In the second diagram photon is emitted from the $p_1$ fermionic line.}
\label{notatation_plot}
\end{figure}

The schematic diagram of the Drell-Yan\,$+$\,jet process is show in Fig.~\ref{notatation_plot}.
We denote the proton projectiles four-momenta as $P_1$ and $P_2$, and $\sp=(P_1+P_2)^2$ is the invariant Mandelstam variable.
We apply the standard Sudakov decomposition of four momenta, e.g. for the DY virtual photon $\gamma^*$ we have 
\beq
q = \alpha_q P_1 + \beta_q P_2 + q_\perp ,
\eeq
with the transverse momentum $q_\perp \cdot P_1=q_\perp \cdot P_2=0$.
The photon virtuality $q^2 = M^2 > 0$ is also the lepton pair invariant mass squared. 
In the light cone coordinates we have $ P_1= (\sqrt{S},0, \vec 0)$ and $ P_2= (0,\sqrt{S}, \vec 0)$, and for any four vectors $u$ and $v$ their scalar product is given by
\beq
u\cdot v=\half ( u^+ v^- + u^- v^+) - \vec{u}_\perp \vec{v}_\perp.
\eeq
Thus,  the transverse part of any four-vector is perpendicular to the collision axis with such a choice of the coordinates.

We treat the initial state partons as collinear and their four momenta are
\beq
p_1= ( x_1 P_1^+, 0, \vec{0}_\perp )\,,~~~~~~~~~~~~~~~p_2= (0, x_2 P_2^-, 0, \vec{0}_\perp) ,
\eeq
where $ P_1^+= P_2^-=\sqrt{S}$.
We additionally define the longitudinal momentum fraction of the fast quark $p_1$ taken by the virtual photon $\gamma^*$,
\beq
z=\frac{q^+}{p_1^+} .
\eeq
The virtual photon and jet momenta are the following
\be
q= \left( x_1 z P_1^+,\, \frac{M^2+q_\perp^2}{ x_1 z P_1^+}, \,\vec{q}_\perp \right),~~~~~~~~~~~~~
p_J= \left( \frac{p_{J\perp}^2}{x_2 P_2^-} ,\, x_2 P_2^-, \,\vec{p}_{J\perp} \right)
\ee
and their rapidities are given by
\begin{align}
y_\gamma &= \frac{1}{2} \ln \frac{q^+}{q^-}= \ln\! \left( \frac{z x_1 \sqrt{S}}{\sqrt{M^2 + q^{2}_\perp}} \right), 
\label{jet_Rapidity}
\\
y_J &= \frac{1}{2} \ln \frac{p_J^+}{p_J^-}= \ln\! \left(\frac{p_{J\perp}}{x_2 \sqrt{S}} \right), 
\end{align}
where $q_\perp=|\vec{q}_\perp|$ and $p_{J\perp}=|\vec{p}_{J\perp}|$.
Defining rapidity difference between photon and jet, 
\be
\Delta Y_{\gamma J}= y_\gamma-y_J\,,
\ee
we find from the above relations
\beq\label{eq:2.8}
z= \frac{ p_{J\perp}\sqrt{M^2 + q_\perp^2} } { x_1 x_2 S}\, {\rm e}^{\Delta Y_{\gamma J}}.
\eeq
The spectral condition $0<z<1$ sets a constraint on allowed values of kinematic variables.

In analogy to the Mueller--Navelet (MN) process \cite{Kwiecinski:2001nh}, we define the rapidity difference between the measured jet and the separated in rapidity the photon plus quark system
\beq\label{eq:2.10}
\Delta Y_{P}=\ln \left( \frac { x_2}{x_g}\right)= \ln \left( \frac {z(1-z) x_1 x_2 S}{ M^2(1-z)+q_\perp^2+
z(k_{1\perp}^2-2\vec k_{1\perp}\, \vec q_\perp)} \right),
\eeq
where $x_g$ is the momentum fraction of the uppermost gluon in the BFKL ladder, see Fig.~\ref{notatation_plot}, which value is fixed by kinematics.
This rapidity difference is an argument of the BFKL kernel, discussed in the next section, while $\Delta Y_{\gamma J}$ is a measurable quantity. The functional dependence of $\Delta Y_{P}$ on $\Delta Y_{\gamma J}$ is obtained by substituting eq.~(\ref{eq:2.8}) to eq.~(\ref{eq:2.10}).

Finally, we introduce the variable
\beq
\rho=\ln \left( \frac{k_{1\perp}^2}{ k_{2\perp}^2} \right),
\eeq
which is built from transverse momenta of the first and the upper most gluon in the BFKL ladder. 
Taking into account that at the jet vertex the transverse momentum
of the initial partons equals zero, we have
\beq
\vec k_{2\perp} = -\vec p_{J\perp}\,.
\eeq
Thus, we always replace $\vec k_{2\perp}$ by the jet transverse momentum $-\vec p_{J\perp}$ in what follows.

Since the photon is virtual, it has three polarizations  which we denote by ${\sigma=0,\pm}$. 

\section{Drell-Yan plus jet cross section}
\label{sec:3}

In the standard approach to the inclusive DY process (where only two leptons are measured) one factorizes leptonic and hadronic degrees of freedom~\cite{Lam:1978pu} and the cross-section is written as an angular distribution of one lepton\footnote{By convention we choose lepton with positive charge.} in the lepton pair center-of-mass frame:
\begin{eqnarray}
\frac{d\sigma^{\textrm{DY,inc}}}{d^4 \, q \, d \Omega } & = & \frac{\alphaEM^2 }{2\, (2\pi)^4 M^4} \left[ (1-\cos ^2 \theta) W^{(L)}_\textrm{inc} + (1+\cos ^2 \theta) W^{(T)}_\textrm{inc} + \right. \nonumber\\
& + & \left.(\sin^2\theta \cos 2\phi)W^{(TT)}_\textrm{inc} + (\sin2\theta \cos \phi) W^{(LT)}_\textrm{inc} \right].
\label{sigAsWcomb}
\end{eqnarray}
In the above expression $\Omega=(\theta,\phi)$ is a solid angle of a positive charged lepton and $q$ is a four-momentum of virtual photon. The coefficients $ W^{(\lambda)}$ with  $\lambda = T, L, TT ,LT$
are called helicity structure functions and do not depend on $\Omega$. They are obtained as appropriate projections of hadronic tensor on the helicity state vectors $\epsilon^\mu_{\sigma=0,\pm}$ of the virtual photon.

For the DY+jet process, where both photon and jet are measured, the structure of (\ref{sigAsWcomb}) is preserved and one can separate leptonic and hadronic degrees of freedom  in the cross section
\begin{eqnarray}
\label{sigAsWcomb_DYj}
\frac{d\sigma^{\textrm{DY+j}}}{ d\Pi d\Omega}  &=& (1-\cos ^2 \theta)  \frac{d\sigma^{(L)}}{d\Pi}+ (1+\cos ^2 \theta)  \frac{d\sigma^{(T)}}{d\Pi} +\nonumber \\
&+& (\sin^2\theta \cos 2\phi)\frac{d\sigma^{(TT)}}{d\Pi}+ (\sin2\theta \cos \phi) \frac{d\sigma^{(LT)}}{d\Pi},
\end{eqnarray}
where $d\Pi$ is the phase space element of a set of kinematic variables for the DY+jet final state: 
\be
d\Pi= dM^2 \,d^2\vec q_\perp\, d^2\vec p_{J \perp} \,d\Delta Y_{\gamma J}.
\ee
The coefficients ${d\sigma^{(\lambda)}}/{d\Pi}$ play the role of structure functions and for convenience we include the normalization factors related to the above choice of the variables into them to write 
\begin{align}\nonumber
\label{sig_master_formula}
\frac{d\sigma^{(\lambda)}}{ dM^2 \,d\Delta Y_{\gamma J} \,d^2 q_\perp \, d^2 p_{J \perp}} & =
\frac{4 \alphaEM^2\alpha_s^2}{(2\pi)^4} 
\int_0^1 dx_1 \int_0^1 d x_2\,\theta(1-z)\,f_{q}(x_1,\mu) f_{\textrm{eff}}(x_2,\mu) \,\times
\\
&\times \frac{1}{M^2 p_{J\perp}^2}\int \frac{d^2 k_{1\perp}}{ k_{1\perp}^2 } \, \Phi^{(\lambda)} ( \vec q_\perp, \vec k_{1\perp}, z)\,
K(\vec{k}_{1\perp}, \vec{k}_{2\perp}=-\vec p_{J\perp}, \Delta Y_{P}),
\end{align}
where the rapidity difference $\Delta Y_P$ is given by eq.~(\ref{eq:2.10}) while $z$ is given by eq.~(\ref{eq:2.8}).
The theta function in the above restricts $z$ to the interval $(0,1)$.
The quantities
\begin{align}\label{f_q_def}
f_{q}(x_1,\mu)  &=\sum_{i=1}^5 e_i^2\left\{f_i (x_1,\mu)+\bar{f}_{i} (x_1,\mu)\right\},
\\
\label{f_eff_def}
f_{\textrm{eff}}(x_2,\mu)  &= f_g(x_2,\mu)+ \frac{C_F}{C_A}\sum_{i=1}^5 \left\{f_i(x_2,\mu) +\bar{f}_{i} (x_2,\mu)\right\}
\end{align}
are built of the collinear parton distribution functions (PDFs) with five quark flavours and gluon, taken at the scale equal to the transverse
mass of   the virtual photon
\be\label{mperp}
\mu= M_\perp\equiv \sqrt{M^2+q_\perp^2}.
\ee
The DY impact factors $\Phi^{(\lambda)}$ for the Gottfried-Jackson 
helicity frame were obtained in \cite{Motyka:2014lya} and they are given by
\begin{align}
\label{eq:impact_factorL}
\Phi^{(L)} ( \vec{q}_\perp, \vec{k}_\perp, z) &= 2\left[ \frac{M(1-z) }{D_1} - \frac{M(1-z) }{D_2 } \right]^2 ,
\\
\Phi^{(T)} (\vec{q}_\perp, \vec{k}_\perp, z) &= \frac{1+(1-z)^2}{2} 
\left[ \frac{\vec{q}_\perp}{D_1}
-\frac{\vec{q}_\perp-z\vec{k}_\perp}{D_2} \right]^2,
\\
\Phi^{(TT)}(\vec{q}_\perp, \vec{k}_\perp, z) &=(1-z)\! \left\{ 
\left[\left(\frac{\vec q_{\perp}}{D_1}- \frac{\vec q_{\perp}-z \vec k_{\perp}}{D_2}\right)\cdot \vec e_x\right]^2
-
\left[\left(\frac{\vec q_{\perp}}{D_1}- \frac{\vec q_{\perp}-z \vec k_{\perp}}{D_2}\right)\cdot \vec e_y\right]^2
\right\},
\\ 
\Phi^{(LT)}(\vec{q}_\perp, \vec{k}_\perp, z) &=(2-z)
\left[ \frac{M(1-z) }{D_1} - \frac{M(1-z) }{D_2 } \right]  
\left[ \frac{\vec q_{\perp}}{D_1}- \frac{\vec q_{\perp}-z \vec k_{\perp}}{D_2}\right] \!\cdot \vec{e}_x,
\label{eq:impact_factorLT}
\end{align}
where $\vec e_x$ and $\vec e_y$ are two orthogonal unit vectors in the transverse plane perpendicular to the collisions axis, and the denominators read
\be
D_1=M^2(1-z)+\vec{q}_\perp^{\ 2}\,,~~~~~~~~~~~~~~~~
D_2 = M^2(1-z)+(\vec{q}_\perp-z\vec{k}_\perp)^2. 
\ee

Notice that in the Gottfried-Jackson helicity frame, the  $\hat x$ polarization axis viewed in the LAB frame  has
the  transverse part always parallel to the transverse momentum of the virtual photon $\qperp$ in this frame. Thus,
we set  $\vec{e}_x \parallel \vec q_\perp$ and in  consequence $\vec q_\perp=(|\vec q_\perp|,0)$.

\subsection{BFKL kernel}

In eq.~(\ref{sig_master_formula}), $K(\vec{k}_{1\perp}, \vec{k}_{2\perp},\Delta Y_P)$ is the BFKL kernel, given by the Fourier decomposition
\beq
K(\vec k_{1\perp},\vec{k}_{2\perp},\Delta Y_P) =\frac{2}{(2\pi)^2|\vec k_{1\perp}||\vec k_{2\perp}|} \left( I_0(\Delta Y_P,\rho)+ \sum_{m=1}^{\infty} 2\cos(m \phi) I_m(\Delta Y_P,\rho) \right),
\label{Fourier_exp_K}
\eeq
where $\phi$ is the azimuthal angle between the transverse momenta  $\vec k_{1\perp}$ and $\vec k_{2\perp}= -\vec p_{J\perp}$
of the exchanged gluons, see Fig.~\ref{notatation_plot}.
We use the solution to BFKL equation, specified by the coefficients $I_m$:
\beq
I_m(\Delta Y_P,\rho) = \int_0^{\infty} d\nu\, R_m^{A}(\nu)\exp(\omega_m^{A}(\nu)\Delta Y_P)\cos(\rho\nu).
\label{BFKL_LO_integral}
\eeq

We consider two cases: the leading logarithmic (LL) approximation and the LL approximation supplemented by a part of the next-to-leading logarithmic corrections in the form of a consistency condition (CC). The BFKL equation with consistency condition was proposed in Refs.\ \cite{Andersson:1995jt,Kwiecinski:1996td}, and later it was found to resum to all orders the leading collinear and anti-collinear corrections to the BFKL kernel \cite{Salam:1998tj,Ciafaloni:1999yw,Salam:1999cn}.
\begin{enumerate}
\item In the LL approximation the BFKL equation solution reads
\beq
\omega^{\rm LL}_m(\nu)=\chi_m(0,\nu)= \as
\left[ 2 \psi(1) - 
\psi \left( {m+1\over 2}+ i \nu \right) - 
\psi \left( {m+1\over 2}- i \nu \right) 
\right],
\label{omega_LL}
\eeq
where $\psi(z)=\Gamma^\prime(z)/\Gamma(z)$ is the digamma function and
\be
R_m^{\rm LL}(\nu)=1.
\ee
\item The solution of the BFKL equation with CC, and with the symmetric choice of the scale, is given by \cite{Kwiecinski:1996td,Salam:1998tj,Ciafaloni:1999yw,Salam:1999cn}
\beq
R_m^{\rm CC}(\nu) = \left[ 1 - 
{d\chi_m(\omega,\nu) \over d\omega} 
\Big|_{\omega=\omega_m^{\rm CC}(\nu)} \
\right]^{-1},
\eeq
where $\omega_m^{\rm CC}(\nu)$ is a solution of the equation
\beq
\omega_m^{\rm CC}(\nu) = \chi_m(\omega_m^{\rm CC}(\nu),\nu),
\label{omega_CC}
\eeq
with the modified BFKL characteristic function 
\beq
\chi_m(\omega,\nu) = \as
\left[ 2 \psi(1) - 
\psi \left( {m+\omega+1\over 2}+ i \nu \right) - 
\psi \left( {m+\omega+1\over 2}- i \nu \right) 
\right].
\label{chi_BFKLdef}
\eeq
\end{enumerate}

\begin{figure}[t]
\begin{center}
\includegraphics[width=.55\textwidth]{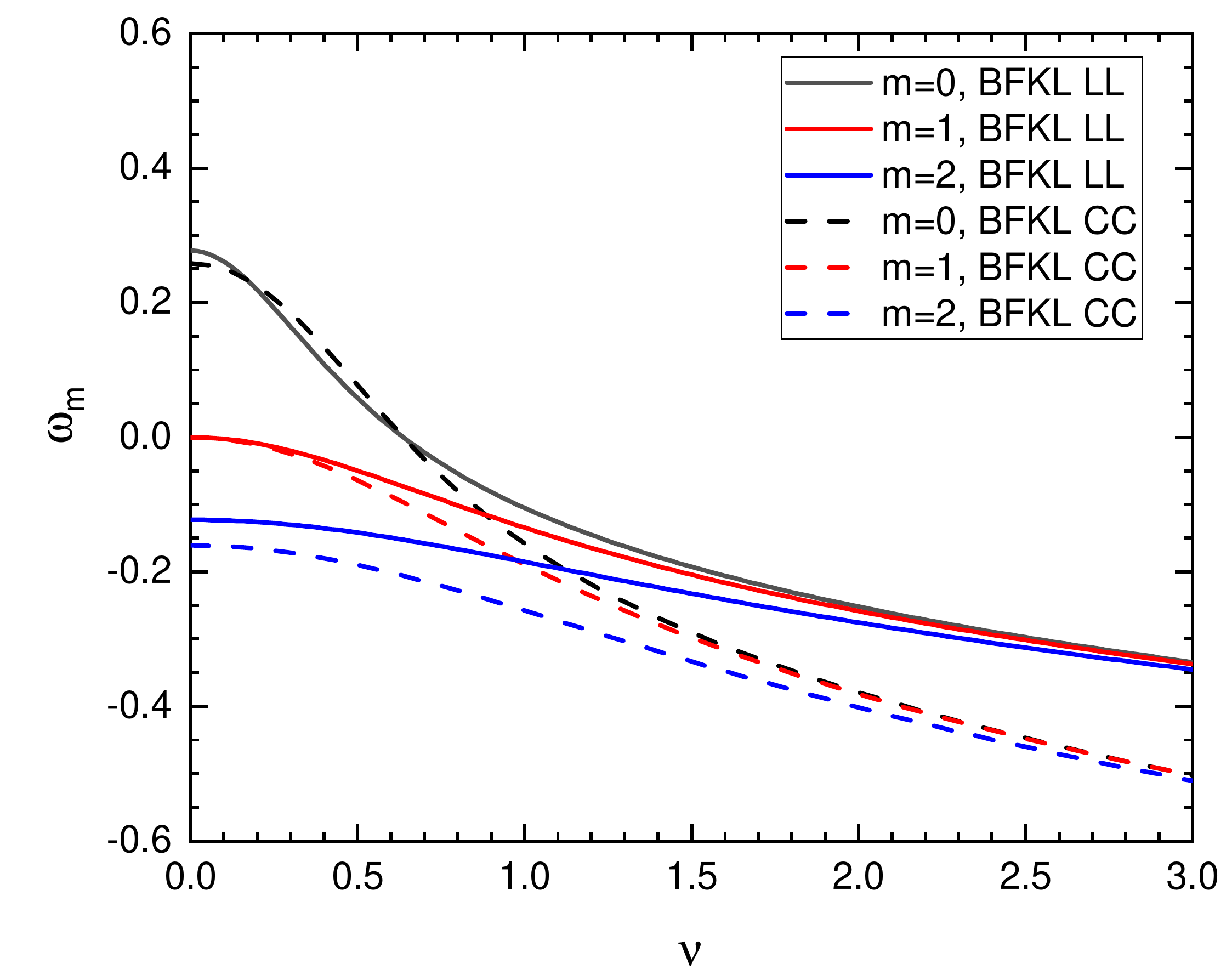}
\end{center}
\caption{The functions $\omega_m^{\rm LL}(\nu)$ (solid lines) and $\omega_m^{\rm CC}(\nu)$ (dashed lines) 
defined by eqs.~(\ref{omega_LL}) and (\ref{omega_CC}), respectively, for $m=0,1,2$.}
\label{omega_nu_plot}
\end{figure}

In Fig.~\ref{omega_nu_plot} we show the functions $\omega_m^{\rm LL}(\nu)$ and $\omega_m^{\rm CC}(\nu)$ for $m=0,1,2$ obtained for the LL (solid lines) and CC (dashed lines) BFKL solution, respectively. 
We choose the values of $\bar{\alpha}_s$ such that the intercept values, $\omega_0^{\rm LL}(0)$ and $\omega_0^{\rm CC}(0)$, are both close to 
the value $0.27$ which allows to successfully describe the HERA data on $F_2$: $\bar{\alpha}_s=0.1$ for LL and $\bar{\alpha}_s=0.15$ for CC. With such a choice, the LL and CC functions for $m=0$ are 
very close to each other up to $\nu\approx 1$, see Fig.~\ref{omega_nu_plot}, which is a dominant region for the integration over $\nu$ 
in eq.~(\ref{BFKL_LO_integral}). The same is true for $m=1$, in which case the two functions equal zero for $\nu=0$ by definition. These two contribution practically
dominate the sum over $m$ in the BFKL kernel (\ref{Fourier_exp_K}). This explains why the numerical results on angular decorrelations, 
presented in Section~\ref{sec:6}, are very similar for the LL and CC BFKL kernels .

In the forthcoming analysis we will also consider the BFKL kernel in the leading order (LO)-Born approximation in which only two gluons in the color singlet state are
exchanged. In this case the exchange kernel reads, 
\be\label{eq:3.13}
K(\vec k_{1\perp},\vec{k}_{2\perp},\Delta Y_P)= \frac{1}{2} \delta^2(\vec k_{1\perp}-\vec{k}_{2\perp}).
\ee


\subsection{Azimuthal angle dependence}

A special attention has to be paid to the azimuthal angle dependence in the transverse plane to the collision axis.
In the LAB frame,  the transverse part of the Gottfried-Jackson $\hat{x}$ polarization axis  is oriented along the positive direction of the photon transverse momentum $\vec q_\perp$, thus the azimuthal angle of the photon $\phi_\gamma=0$.

Therefore, we define the following angles with respect to $\hat x$ 
in the transverse plane  for the jet and the upper most gluon in the BFKL ladder  transverse momenta
\be
\phi_J= \measuredangle(\vec p_{J\perp}, \hat x) \,,~~~~~~~
\phi_g= \measuredangle(\vec k_{1\perp}, \hat x)
\ee
and consider the differences
\be
\phi_{\gamma J} =\pi- \phi_J\,,~~~~~~~~~~~~\phi_{Jg} = \pi-(\phi_J- \phi_g).
\ee
The jet and photon are back-to-back in transverse plane when $\phi_{\gamma J} =0$. The same is true for the jet and the gluon $\vec{k}_{1\perp}$ when $\phi_{Jg}=0$.
For further analysis, we choose as independent  angles $\phi_{\gamma J}$ and $\phi_g$.  The first angle
is an observable while the latter is the integration variable in the integral over $\vec k_{1\perp}$.
Therefore,  we rewrite (\ref{sig_master_formula}) in the form
\begin{align}\nonumber
\label{sig_master_formula_int}
&\frac{d\sigma^{(\lambda)}}{dM d\Delta Y_{\gamma J} dq_\perp \, d p_{J \perp} d\phi_{\gamma J} } = 
\frac{16\, \alphaEM^2 \alpha_s^2}{(2\pi)^5}  \,
\frac{q_\perp}{M p_{J\perp}^2}
\int_0^1 dx_1 \int_0^1 d x_2 \,\theta(1-z)\, f_q(x_1,\mu) f_{\textrm{eff}}(x_2,\mu) \,\times
\\ 
&~~~\times 
\int \frac{d k_{1\perp} }{ k_{1\perp}^2} \int_0^{2\pi} d \phi_g \,  \Phi^{(\lambda)} ( \vec q_\perp, \vec k_{1\perp}, z) \Big\{ I_0(\Delta Y_{P}, \rho)+ \sum_{m=1}^{\infty} 2\cos(m \phi) I_m(\Delta Y_{P}, \rho)
\Big\},
\end{align}
where the angle in the BFKL kernel is given by $\phi=2\pi-(\phi_{\gamma J}+\phi_g)$  and
\be
\cos(m \phi)= \cos\left[m (\phi_{\gamma J}+ \phi_{g} )\right]= \cos(m \phi_{\gamma J}) \cos(m \phi_{g})- \sin(m \phi_{\gamma J})\sin(m \phi_{g}).
\ee
The impact factors $\Phi^{(\lambda)}$ are even with respect to the transformation $\phi_g\to-\phi_g$ and the term 
proportional to $\sin(m \phi_{g})$ vanishes when integrated over $\phi_g$. 
Thus, we obtain the following cross-section for the DY$\,+\,$jet production
\begin{align}\nonumber
\frac{d\sigma^{(\lambda)}}{dM d\Delta Y_{\gamma J} dq_\perp \, dp_{J \perp}d\phi_{\gamma J}} & = 
\mathcal{I}^{(\lambda)}_0(M,\Delta Y_{\gamma J}, q_\perp, p_{J \perp})~+
\\
&+\sum_{m=1}^{\infty} 2\cos(m \phi_{\gamma J})\, \mathcal{I}^{(\lambda)}_m(M,\Delta Y_{\gamma J}, q_\perp, p_{J \perp}),
\label{fourier_exp_of_master_form}
\end{align}
where the Fourier coefficients, for $m=0,1,2 \ldots$, have the form:
\begin{align}\nonumber
\mathcal{I}^{(\lambda)}_m (M,\Delta Y_{\gamma J}, q_\perp, p_{J \perp}) &=
\frac{16\, \alphaEM^2 \alpha_s^2}{(2\pi)^5}  \,
\frac{q_\perp}{M p_{J\perp}^2}
\int_0^1 dx_1 \int_0^1 d x_2\,\theta(1-z)\, f_q(x_1,\mu) f_{\textrm{eff}}(x_2,\mu) \,\times
\\
&\times
\int \frac{d k_{1\perp} }{ k_{1\perp}^2} \int_0^{2\pi} d \phi_g\, \Phi^{(\lambda)} ( \vec q_\perp, \vec k_{1\perp}, z)\cos\left(m\phi_{g}\right) I_m(\Delta Y_{P}, \rho)
\label{Four_coeff_defintion}
\end{align}
In the LO-Born approximation (\ref{eq:3.13}), the DY$\,+\,$jet cross section in the given helicity state reads
\begin{align}\nonumber
\label{sig_master_formula_int_exp_LO}
\frac{d\sigma^{(\lambda)}}{dM d\Delta Y_{\gamma J} dq_\perp \, d p_{J \perp}d\phi_{\gamma J}} & =
\frac{4\, \alphaEM^2 \alpha_s^2}{(2\pi)^3}  \,
\frac{q_\perp}{M p_{J\perp}^3}
 \int_0^1 dx_1 \int_0^1 d x_2 \,  \theta(1-z)\, \times \nonumber 
\\
& \times f_q(x_1,\mu)\, f_{\textrm{eff}}(x_2,\mu)\, \Phi^{(\lambda)} ( \vec q_\perp, -\vec p_{J\perp}, z).
\end{align}

\subsection{Lepton angular distribution coefficients}
\label{sec:4}

The integration of (\ref{sigAsWcomb_DYj}) over the full spherical angle $\Omega$ gives the helicity-inclusive cross section:
\beq
\frac{d\sigma^{\textrm{DY+j}}}{d\Pi } \equiv \int d\Omega \, \frac{d\sigma^{\textrm{DY+j}}}{d\Pi d\Omega}  = \frac{16\pi}{3}\left( \frac{d\sigma^{(T)}}{d\Pi} + \frac{1}{2} \frac{d\sigma^{(L)}}{d\Pi} \right).
\label{hel_incl_xsect}
\eeq
In the inclusive DY process it is useful to define normalized structure functions. We follow this approach and define
for the DY+jet process:
\begin{equation}
A_0 = \frac{d\sigma^{(L)}  }{ d\sigma^{(T)} + d\sigma^{(L)} /2 }, \ \ \ A_1 = \frac{d\sigma^{(LT)}  }{ d\sigma^{(T)} +  d\sigma^{(L)}/2  } , \ \ \ A_2 =  \frac{2d\sigma^{(TT)}  }{ d\sigma^{(T)} +  d\sigma^{(L)}/2  }\, .
\label{A_coeff_def}
\end{equation}

Lam and Tung proved the following relation valid at the LO and NLO for the DY $qg$ channel in the collinear leading twist approximation \cite{Lam:1978pu,Lam:1980uc}:
\beq
 d\sigma^{(L)}- 2 d\sigma^{(TT)}=0~~~~~~~~ \textrm{ or } ~~~~~~~~A_0-A_2=0.
\label{Lam-Tung_combination}
\eeq
As it was shown in \cite{Motyka:2016lta}, the combination $A_0-A_2$ is sensitive to partons' transverse momenta.

The coefficients $d\sigma^{(\lambda)}/d\Pi $ (like the structure functions $W^{(\lambda)}_{\textrm{inc}}$ in the inclusive DY) are computed for a particular choice of the polarization axes, i.e. Gottfried--Jackson frame. Since most of experimental results are provided in the Collins--Soper helicity frame, we apply an additional rotation of our impact factors, see appendix A of ref.~\cite{Motyka:2016lta} for the form of rotation matrix.
 
One can find several combinations of structure functions which are invariant w.r.t.~the change of the helicity frame. Obviously, the helicity-inclusive cross section (\ref{hel_incl_xsect}) is one of them. 
It also turns out that  the Lam--Tung combination (\ref{Lam-Tung_combination}) has this property. More information about the frames used to describe the lepton pair and relation between them can be found in \cite{Faccioli:2010kd}.

\subsection{Mueller-Navelet jets}
\label{sec:5}

For a comparison with the DY+jet results, we present also  formulas for the Mueller-Navelet (MN) jet production.
In this case, the DY form factors in (\ref{sig_master_formula}) 
should be replaced by the jet form factor (which is a delta function in the leading order approximation). 
Additionally, the singlet quark distribution $f_{q}$ should be replaced by the effective distribution $f_{\textrm{eff}}$ given by eq.~(\ref{f_eff_def}). Thus
\beq
\frac{d\sigma^{\textrm{MN}}}{d\Delta Y_{IJ}\, d^2 p_{I \perp} \, d^2 p_{J \perp} }  = \frac{(C_A \alpha_s)^2}{p_{I\perp}^2 p_{J\perp}^2} \int_0^1 dx_1
\ f_{\textrm{eff}}(x_1,\mu) \, x_2 f_{\textrm{eff}}(x_2,\mu) K( \vec p_{I \perp}, -\vec p_{J\perp}, \Delta Y_P),
\label{sig_master_formula_MN}
\eeq
where $\vec p_{I \perp}$ and $\vec p_{J\perp}$ are transverse momenta of the two jets and their rapidities are given by
\be
y_I = \ln \left(\frac{x_1 \sqrt{S}}{p_{I\perp}} \right),~~~~~~~~~~~~~~~~~~~~~
y_J = \ln \left(\frac{p_{J\perp}}{x_2 \sqrt{S}} \right).
\ee
Their difference is equal to
\beq
\Delta Y_{IJ}= y_I-y_J=\ln \left(\frac{x_1 x_2 S}{p_{I\perp}p_{J\perp}} \right).
\eeq
Since $\Delta Y_{IJ}$ is fixed in the MN jet analysis, only one of the two longitudinal momentum fractions of the initial partons 
is an independent variable. Similarly to the pure DY case, we expand the 
BFKL kernel using formula (\ref{Fourier_exp_K}) in which $\phi=\phi_{IJ}=\pi -(\phi_I - \phi_J)$ is the angle between jets' transverse momenta, $\Delta Y_P=\Delta Y_{IJ}$ and 
$\rho=\ln({p_{I\perp}^2}/{ p_{J\perp}^2})$.

\section{Numerical results}
\label{sec:6}

In this section we present numerical results obtained for the LHC hadronic center-of-mass energy, $\sqrt S=13~{\rm TeV}$. 
For the collinear parton distributions which enter $f_{q}$ and $f_{\textrm{eff}}$, we use the NLO MMHT2014 set \cite{Harland-Lang:2014zoa} with the scale $\mu=M_\perp$, see eq.~(\ref{mperp}). 
We also impose the following cuts for the rapidities of the photon and the jet:
\be
|y_\gamma|<4\,,~~~~~~~~~~~~~~~|y_J|<4.7.
\ee


\subsection{Helicity-inclusive DY+jet cross section}

\begin{figure}[t]
\begin{center}
\includegraphics[width=.45\textwidth]{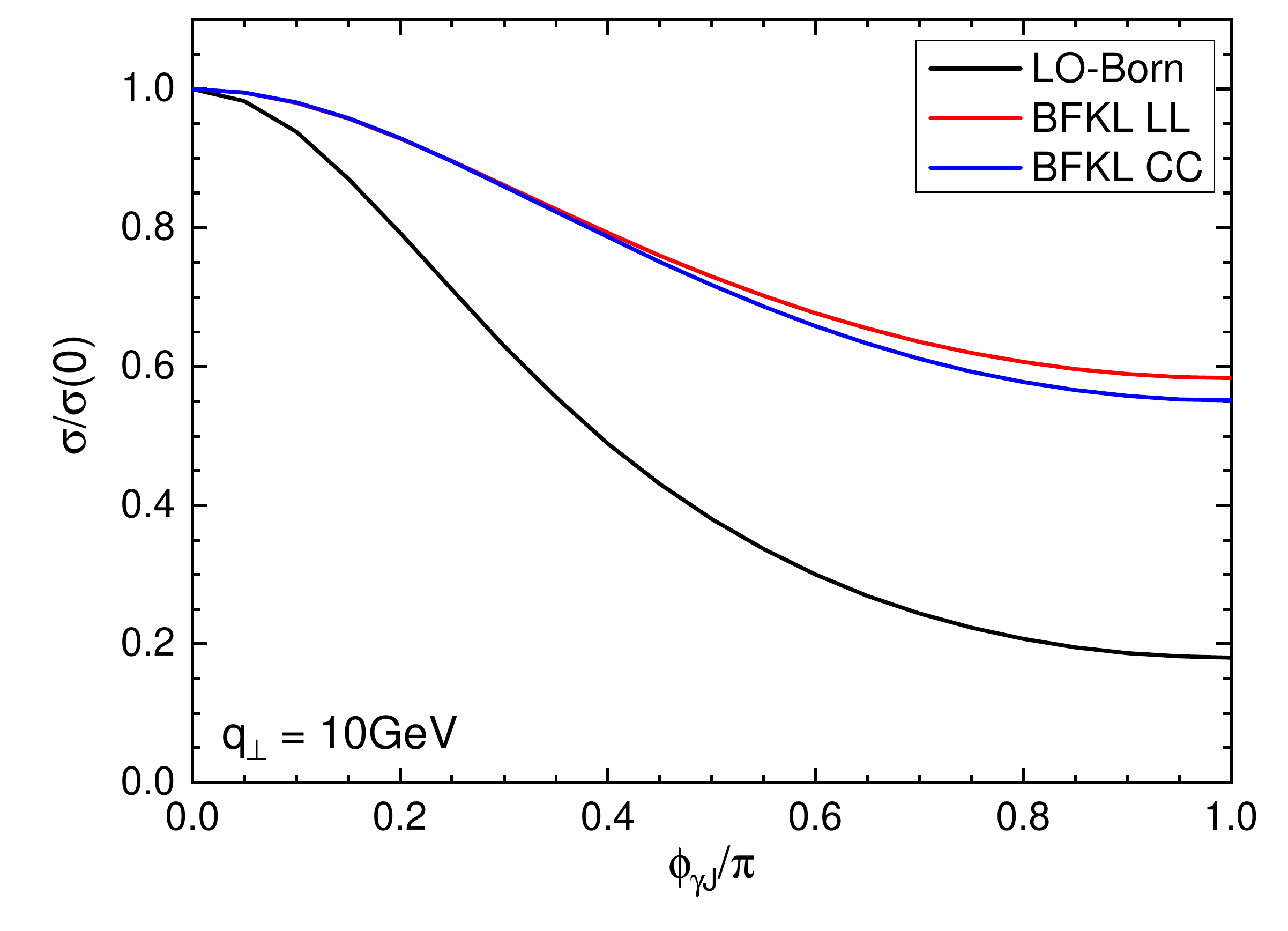}
\includegraphics[width=.45\textwidth]{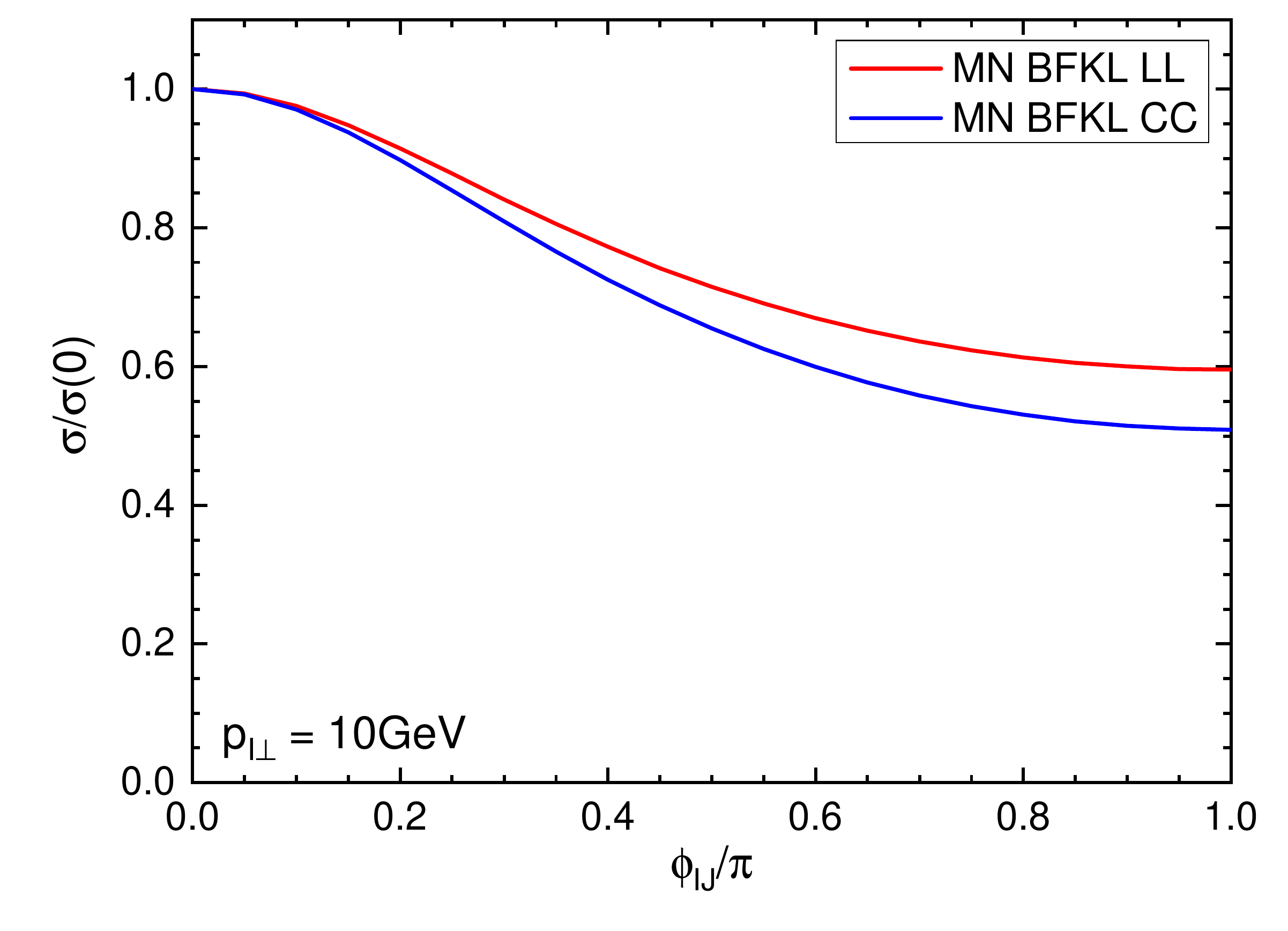}
\includegraphics[width=.45\textwidth]{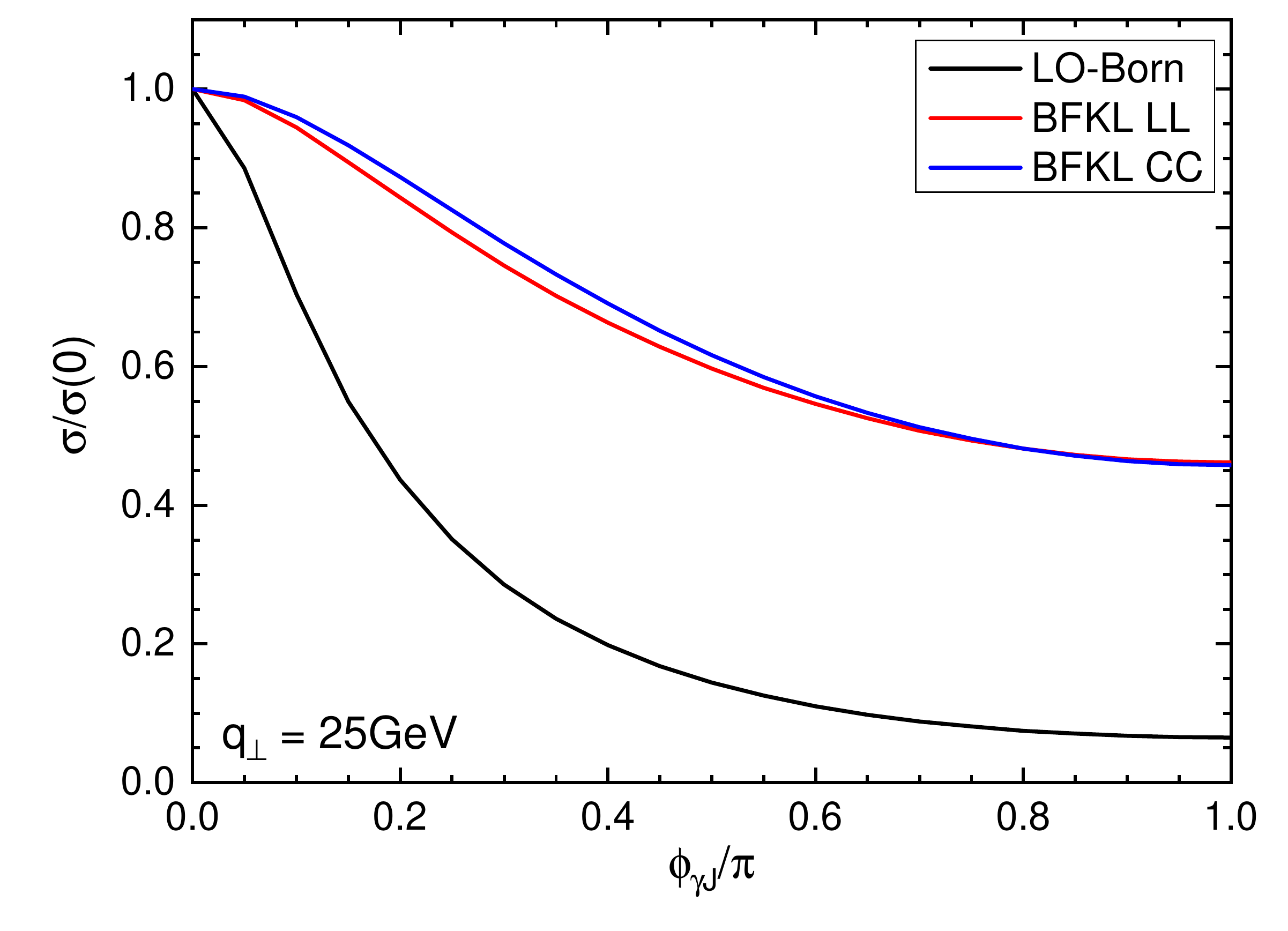}
\includegraphics[width=.45\textwidth]{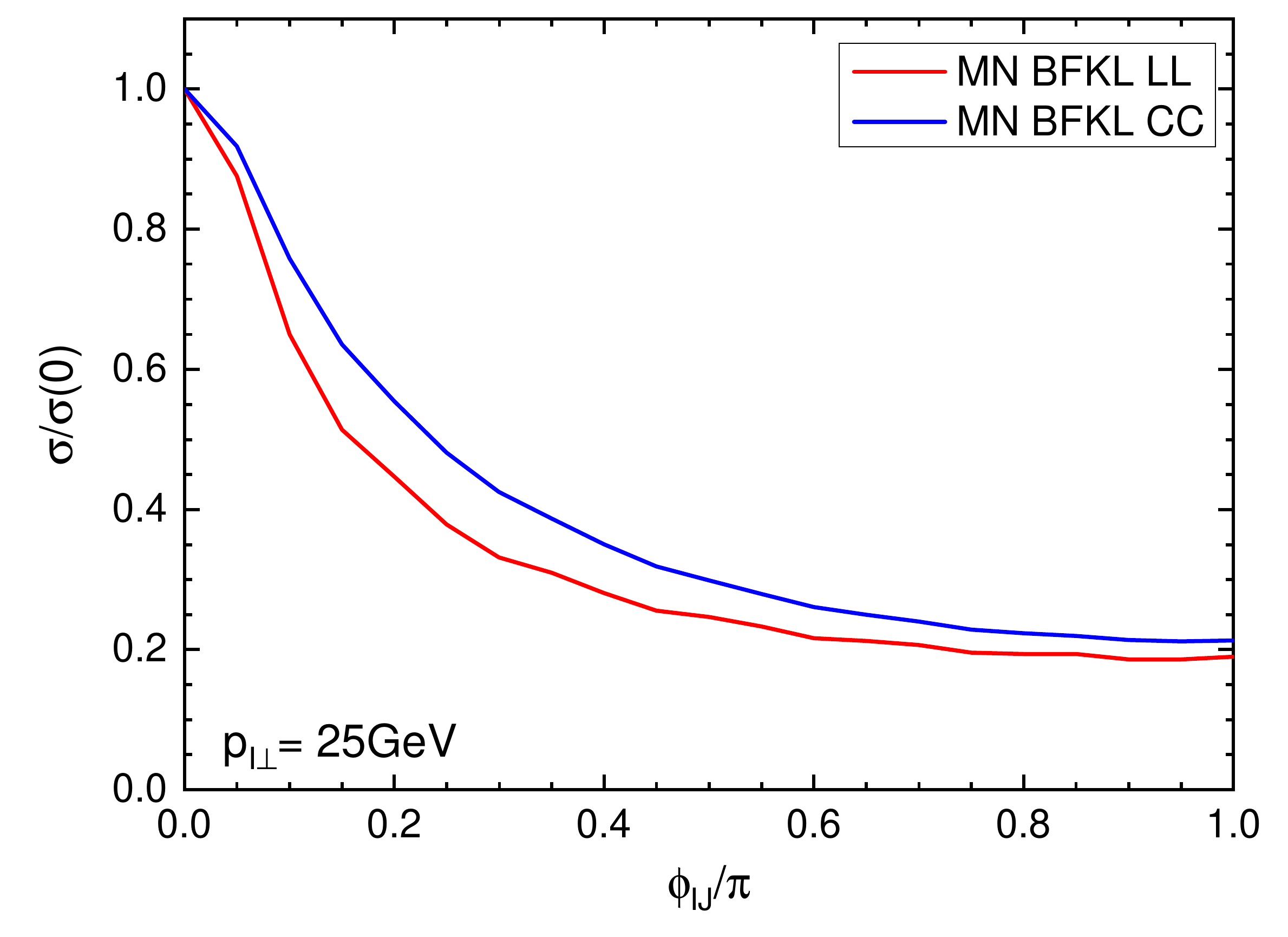}
\includegraphics[width=.45\textwidth]{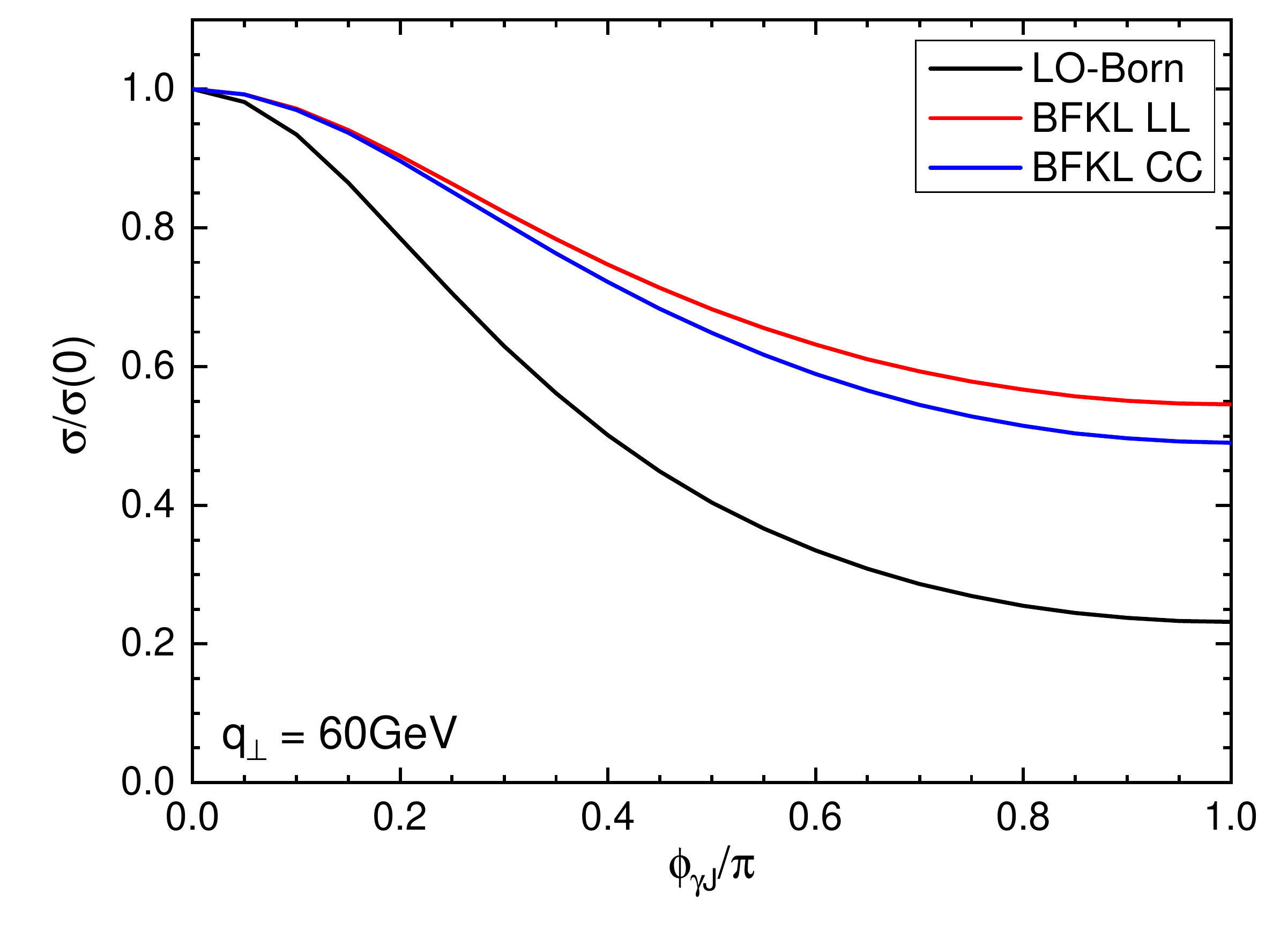}
\includegraphics[width=.45\textwidth]{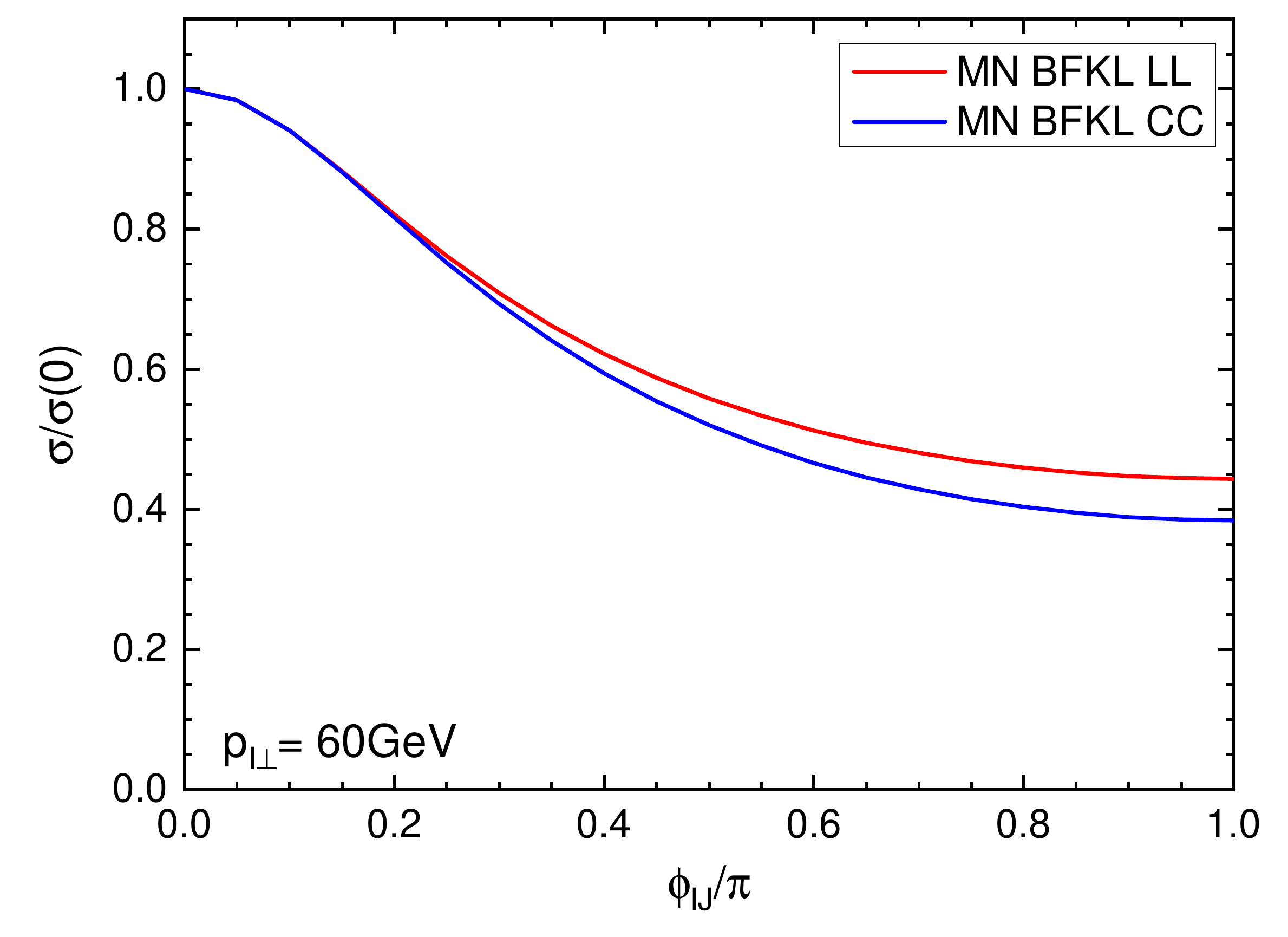}
\end{center}
\caption{The dependence on the azimuthal angle of the normalized helicity-inclusive cross section for the DY+jet (left column) and Mueller-Navelet jets (right column) productions . The following values of parameters are used: 
$p_{J\perp}=30~{\rm GeV}$, $\Delta Y_{\gamma J}=\Delta Y_{I J}=7$ and $M=35~{\rm GeV}$. 
The LO Mueller-Navelet distribution is not shown since $d\sigma^{MN}|_{LO}=0$ when $p_{I\perp}\ne p_{J\perp}$, see (\ref{sigmaMN_deltaDir}). Angles $\phi_{\gamma J}$ and $\phi_{I J}$ are defined such that they equal zero for configurations back-to-back.
}
\label{tot_xsect_BFKLvsLO}
\end{figure}

We start by showing in Fig.~\ref{tot_xsect_BFKLvsLO} (left column) the normalized helicity-inclusive cross section (\ref{hel_incl_xsect})
\be
\frac{d\sigma^{\textrm{DY+j}}(\phi_{\gamma J})}{d\sigma^{\textrm{DY+j}}(0)}
=
\left(\frac{d\sigma^{(T)}(\phi_{\gamma J})}{d\Pi}+ \frac{1}{2}\frac{d\sigma^{(L)}(\phi_{\gamma J})}{d\Pi}\right) \Bigg/
\left(\frac{d\sigma^{(T)}(0)}{d\Pi}+ \frac{1}{2}\frac{d\sigma^{(L)}(0)}{d\Pi}\right)
\label{hel_incl_xsect_norm}
\ee
as a function of the azimuthal jet-photon angle $\phi_{\gamma J}$ for fixed values of $M,\Delta Y_{\gamma J},q_\perp$ and $p_J$. We computed this ratio for 
the three cases of the BFKL equation treatment, discussed in Section~\ref{sec:3}: the leading order LO-Born approximation, and the LL and CC approximations.

As expected, the BFKL gluon emissions lead to a strong decorrelation in the azimuthal angle in comparison to the LO-Born case. This effect does not depend on the value of the photon transverse momentum $\qperp$, which we illustrate by showing the angular dependence for $\qperp=10,\,25$ and $60~{\rm GeV}$. 
We observe that the two considered BFKL models with $\bar{\alpha}_s$ adjusted to the $F_2$ HERA data lead to
similar predictions on the normalized azimuthal dependence. Nevertheless, the BFKL model with CC is more realistic since it resums to all orders the collinear and anti-collinear double logarithmic corrections \cite{Salam:1998tj,Ciafaloni:1999yw,Salam:1999cn}.

In Fig.~\ref{tot_xsect_BFKLvsLO} we also compare the angular decorrelation for the DY$\,+\,$jet (left) and MN jet (right) productions for the same values of the jet
and the photon transverse momenta, $\qperp=p_{I\perp}$, and the rapidity difference $\Delta Y_{\gamma J}=\Delta Y_{IJ}$. We see that the photon decorrelation is stronger in comparison to the MN jet process, which is what we expected due to the more complicated final state with one more particle. However, looking from a pure theoretical side, the differences between the cases with the BFKL emissions and the Born calculations is stronger in the MN case. In the latter case, there is no decorrelation and the two jets are produced back-to-back in the LO-Born approximation,
\beq
\label{sigmaMN_deltaDir}
\frac{d\sigma^{MN}}{d^2 p_{I \perp} \, d^2 p_{J \perp}}\Bigg|_{LO} \sim ~\delta^2( \vec{p}_{I \perp}+ \vec{p}_{J \perp}).
\eeq
In the DY$\,+\,$jet system we are dealing with a three particle final state in the LO-Born approximation, i.e. two jets and a photon, and the Dirac delta is smeared out. 

\begin{figure}
\begin{center}
\includegraphics[width=.46\textwidth]{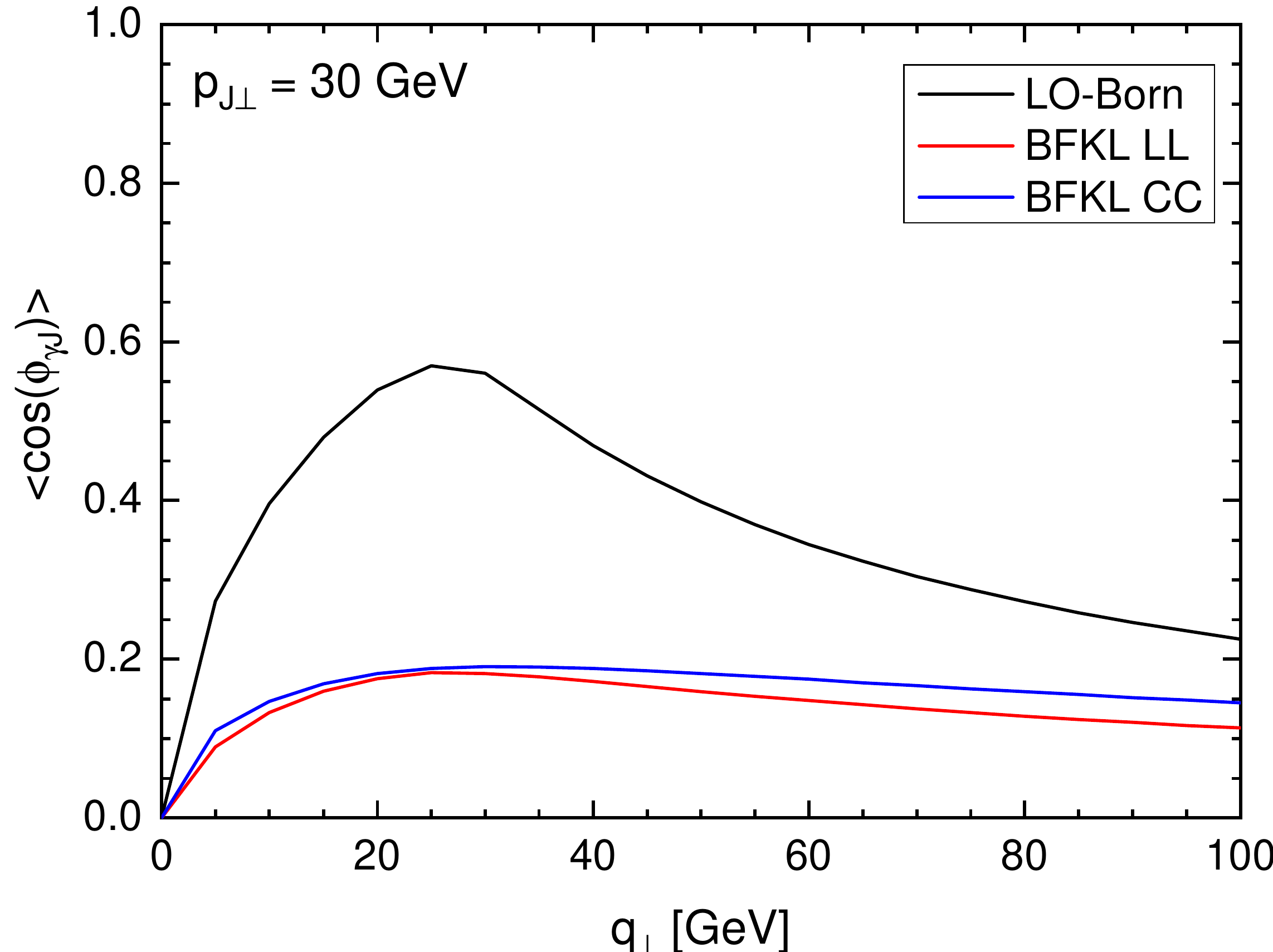}
\includegraphics[width=.46\textwidth]{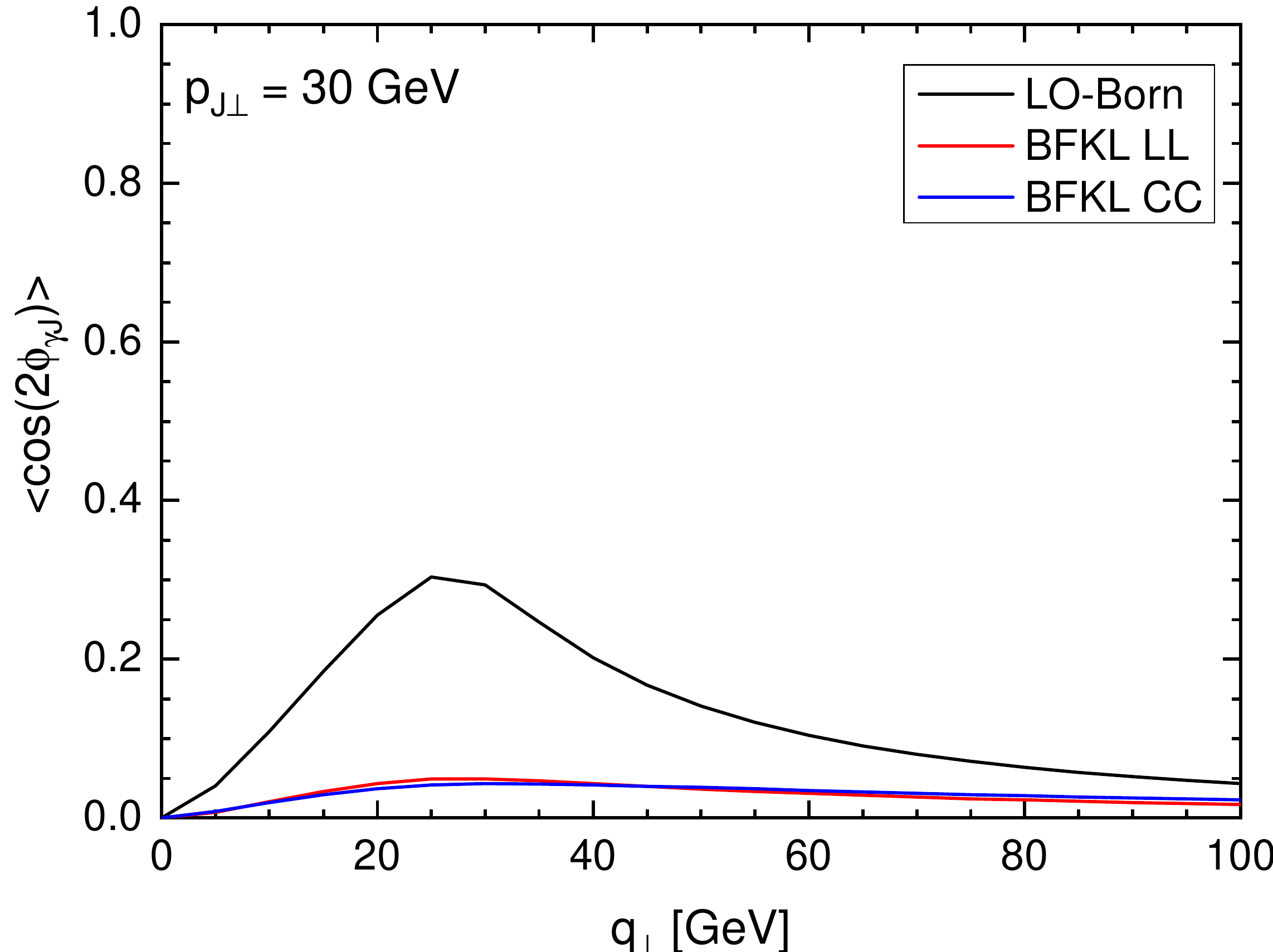}
\includegraphics[width=.46\textwidth]{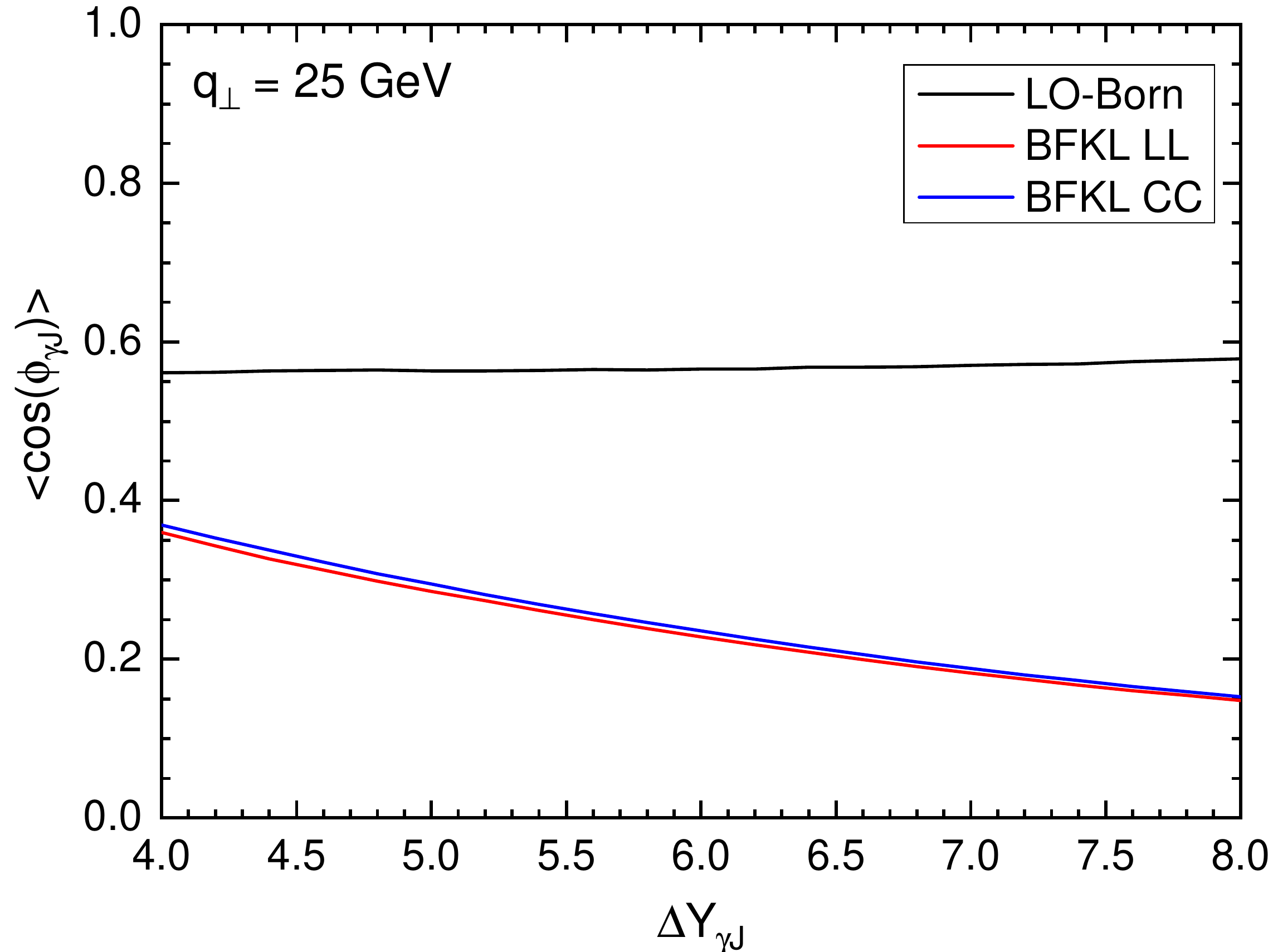}
\includegraphics[width=.46\textwidth]{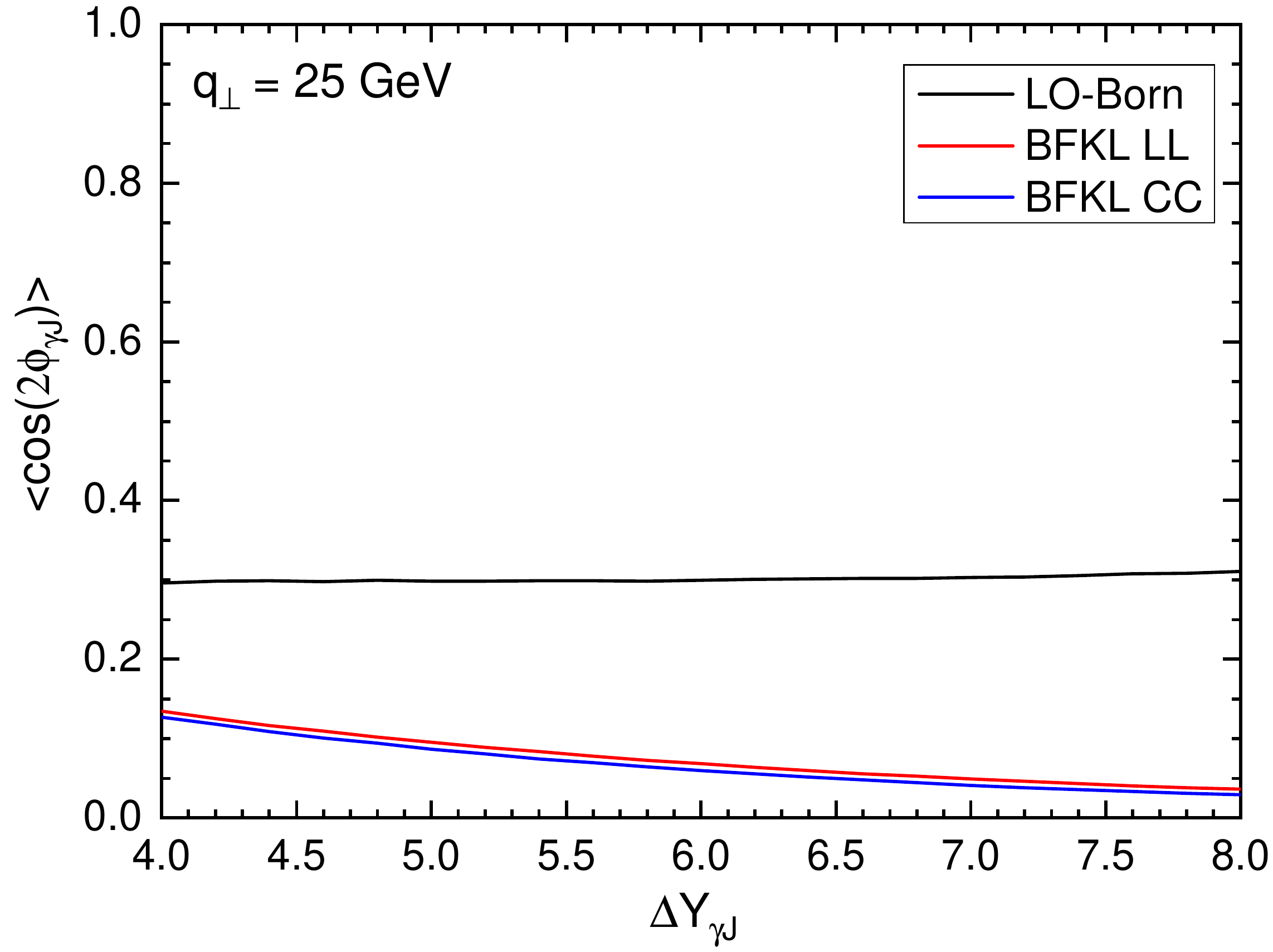}
\end{center}
\caption{
The mean cosine $\langle \cos (n\phi_{\gamma J})\rangle$ for $n=1$ (left) and $n=2$ (right) as a function of
the photon momentum $q_\perp$ (upper row) and the photon-jet rapidity difference $\Delta Y_{\gamma J}$ (lower row).
We choose $\Delta Y_{\gamma J}=7$ for the upper plots and $q_\perp=25~{\rm GeV}$ for the lower plots.
The jet momentum $p_{J\perp}=30~{\rm GeV}$ and the invariant mass of photon-jet system $M=35~{\rm GeV}$ in all cases. 
}
\label{cos_qT_dep}
\end{figure}

For similar transverse momenta of the probes, the angular decorrelation of BFKL driven cross-sections is much stronger for the associated DY and jet production, than it is for the Mueller--Navelet jets --- see Fig.\ \ref{tot_xsect_BFKLvsLO}, the middle row. This may be understood by inspecting the lowest order contributions to both the processes in this kinematical setup. For the MN jets, the first contribution appears at the ${\cal O}(\alpha_s ^3)$ order, from a $2\to 3$ parton process, i.e.\ when at least one iteration of the BFKL kernel is performed. The additional emission is necessary to move the MN jets out of the back-to-back configuration. On the other hand, if the transverse momenta of the jets have similar values, the transverse momentum of the additional emission tends to be small w.r.t.\ the jet momenta, and hence it does not lead to a strong decorrelation. 
In contrast, in the associated virtual photon and jet production, the lowest order process is already at $2\to 3$ level, (e.g.\ $q + g \to q + g + \gamma^*$), and there occurs some angular decorrelation due to the additional quark jet, before the BFKL emissions are included. This decorrelation is further enhanced by the additional gluon emissions. Hence, while the decorrelation for the DY plus jet production is present already at the lowest order, for the MN jets with similar transverse momenta, it only starts at the NLO as a strongly constrained effect. 

When transverse momenta of the probes are strongly unbalanced  -- see first and last row of Fig.~\ref{tot_xsect_BFKLvsLO} -- the additional emission carries significant transverse momenta w.r.t.\ the jets momenta and this implies larger decorrelation than for the balanced probes. In this case angular decorrelation in the DY+jet process is similar to that for the MN jets: the strong additional emission dilutes the difference between two-particles and three-particles final state.

\subsection{More on azimuthal decorrelations}

In the analysis of the azimuthal decorrelation of the MN jets, the mean values of cosines of the azimuthal angle between jets are useful quantities since they can be measured at experiments with good precision. Thus, we follow the idea to study them and define the following quantity for the DY$\,+\,$jet production
with a given polarization $\lambda$:
\beq
{\langle \cos (n\phi_{\gamma J})\rangle}^{(\lambda)} =\frac{\int_0^{2\pi} d\phi_{\gamma J} \ \frac{d\sigma^{(\lambda)}}{dM d\Delta Y_{\gamma J} dq_\perp \, d p_{J \perp}d\phi_{\gamma J}}\cos( n\phi_{\gamma J}) }
{\int_0^{2\pi} d\phi_{\gamma J} \ \frac{d\sigma^{(\lambda)}}{dM d\Delta Y_{\gamma J} dq_\perp \, d p_{J \perp}d\phi_{\gamma J}} }\, ,
\eeq
where the cross section is given by eq.~(\ref{fourier_exp_of_master_form}) for the LL and CC cases and by eq.~(\ref{sig_master_formula_int_exp_LO})
in the LO-Born approximation. Since the coefficients $\mathcal{I}^{(\lambda)}_m$ in eq.~(\ref{fourier_exp_of_master_form}) do not depend on $\phi_{\gamma J}$, the mean cosine in the BFKL case is given by
\beq
\langle \cos (n\phi_{\gamma J})\rangle=  \frac{ \mathcal{I}^{(T)}_n+\mathcal{I}^{(L)}_n/2}{ \mathcal{I}^{(T)}_0+\mathcal{I}^{(L)}_0/2 }\, ,
\eeq
where we skip the symbol $\lambda$ for the helicity-inclusive production.

In Fig.~\ref{cos_qT_dep} we show the mean $\langle \cos (n\phi_{\gamma J})\rangle$ for $n=1$ and $2$
as a function of the photon transverse momentum $q_\perp$ (upper row plots) for a given value of the jet transverse momentum $p_{J\perp}$ in the three indicated in the plot cases. We see that the values of the mean cosines are much smaller in the LL and CC cases which is an indication of a stronger azimuthal decorrelation in comparison to the LO-Born case. All functions have maximum at $q_\perp\sim p_{J\perp}=30$ GeV. One should expect this behaviour since the strongest back-to-back correlation (the biggest cosine value) is possible when photon's transverse momentum balances the transverse momentum of the jet. Once again, the BFKL emissions in the LL and CC approximations dilute this effect significantly. In the lower row of Fig.~\ref{cos_qT_dep} we show the dependence of the mean cosines on the photon-jet rapidity difference $\Delta Y_{\gamma J}$. As expected, the cosine values in the LL and CC approximations decrease with growing rapidity difference since more BFKL emissions are possible, causing stronger decorrelation. On the other hand, in the LO-Born approximation there are no emissions and the cosine values almost not depend on the rapidity difference.

\begin{figure}
\begin{center}
\includegraphics[width=.46\textwidth]{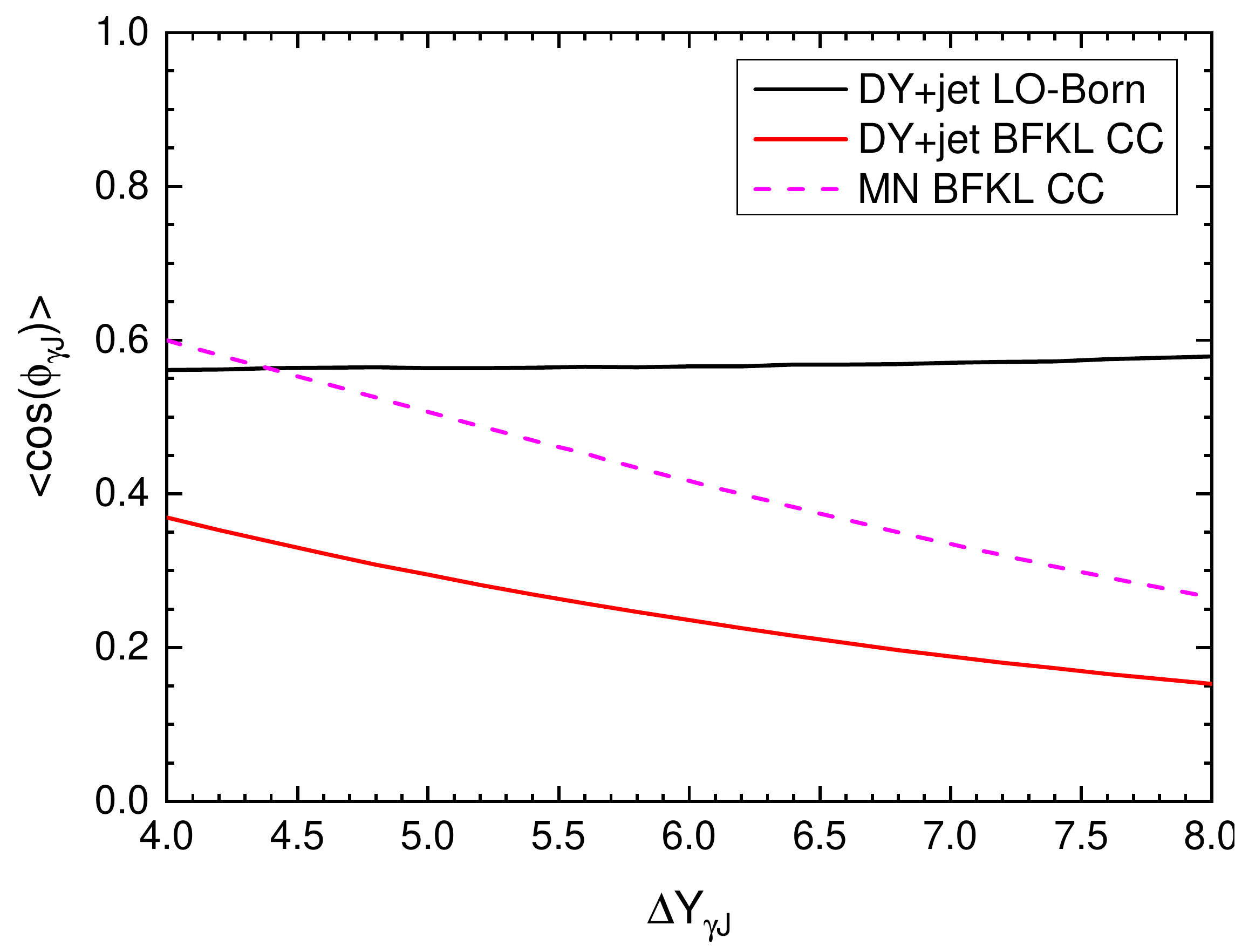}
\includegraphics[width=.46\textwidth]{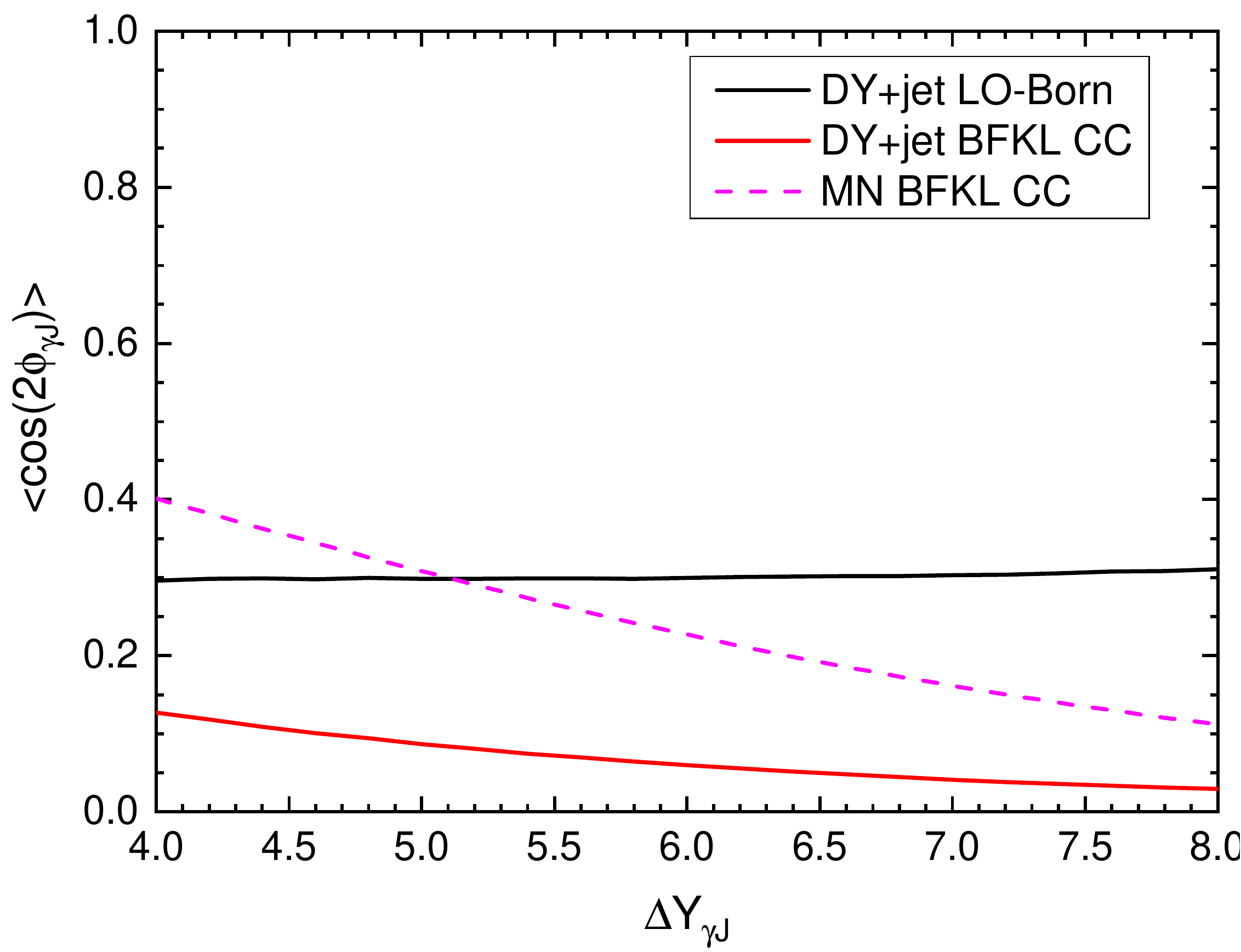}
\end{center}
\caption{The mean cosine of the photon-jet angle $\langle \cos (n\phi_{\gamma J})\rangle$  for $n=1$ (left) and $n=2$ (right) 
for the DY+jet (solid lines) and the MN jets (dashed line) as a function of the rapidity difference $\Delta Y_{\gamma J}$. The parameters are the following 
$q_\perp=p_{I\perp}=25~{\rm GeV}$, $p_{J\perp}=30~{\rm GeV}$ and $M=35~{\rm GeV}$.
}
\label{cos_Y_dep_DYvsMN}
\end{figure}

In Fig.~\ref{cos_Y_dep_DYvsMN} we perform the comparison between the DY+jet (solid lines) and MN jet (dashed lines) processes in terms of the mean cosines 
$\langle \cos (\phi_{\gamma J})\rangle$ for $n=1$ and $n=2$ as a functions of $\gamma$--jet or jet--jet rapidity difference in the indicated on the plots approximations.
In general, we see stronger decorrelations for the DY$\,+\,$jet production that for the MN jet production in both approximations: 
the LO-Born and the BFKL with CC. Note that, the mean cosine values equal one for the LO-Born MN jets 
when both jets have the same transverse momentum. On the other hand, if the jets have different transverse momenta 
(which is the case shown on Fig.~\ref{cos_Y_dep_DYvsMN}), the mean cosine value is not well defined at the Born level. 

\subsection{Angular coefficients of DY leptons}

\begin{figure}
\begin{center}
\includegraphics[width=.46\textwidth]{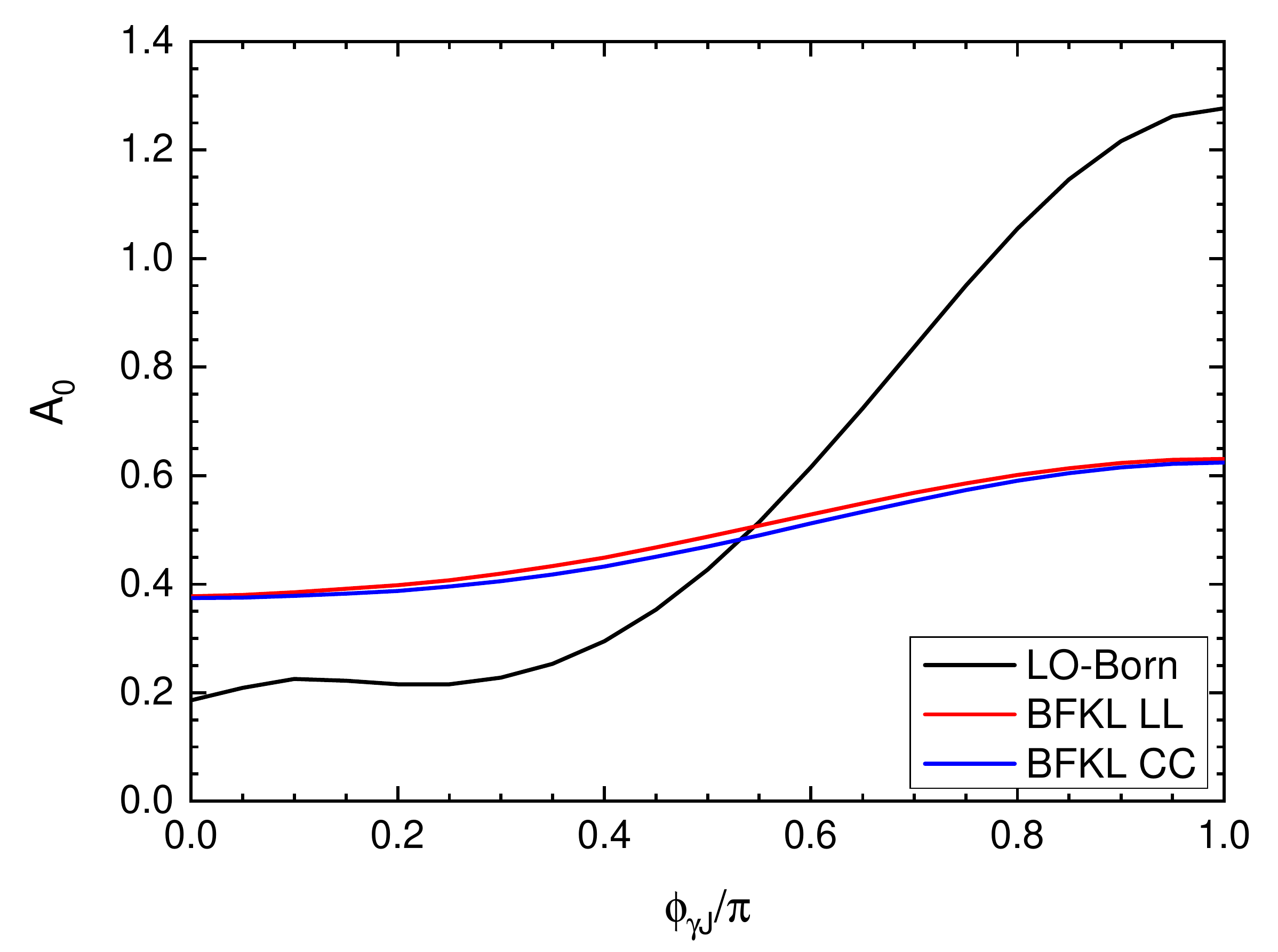}
\includegraphics[width=.46\textwidth]{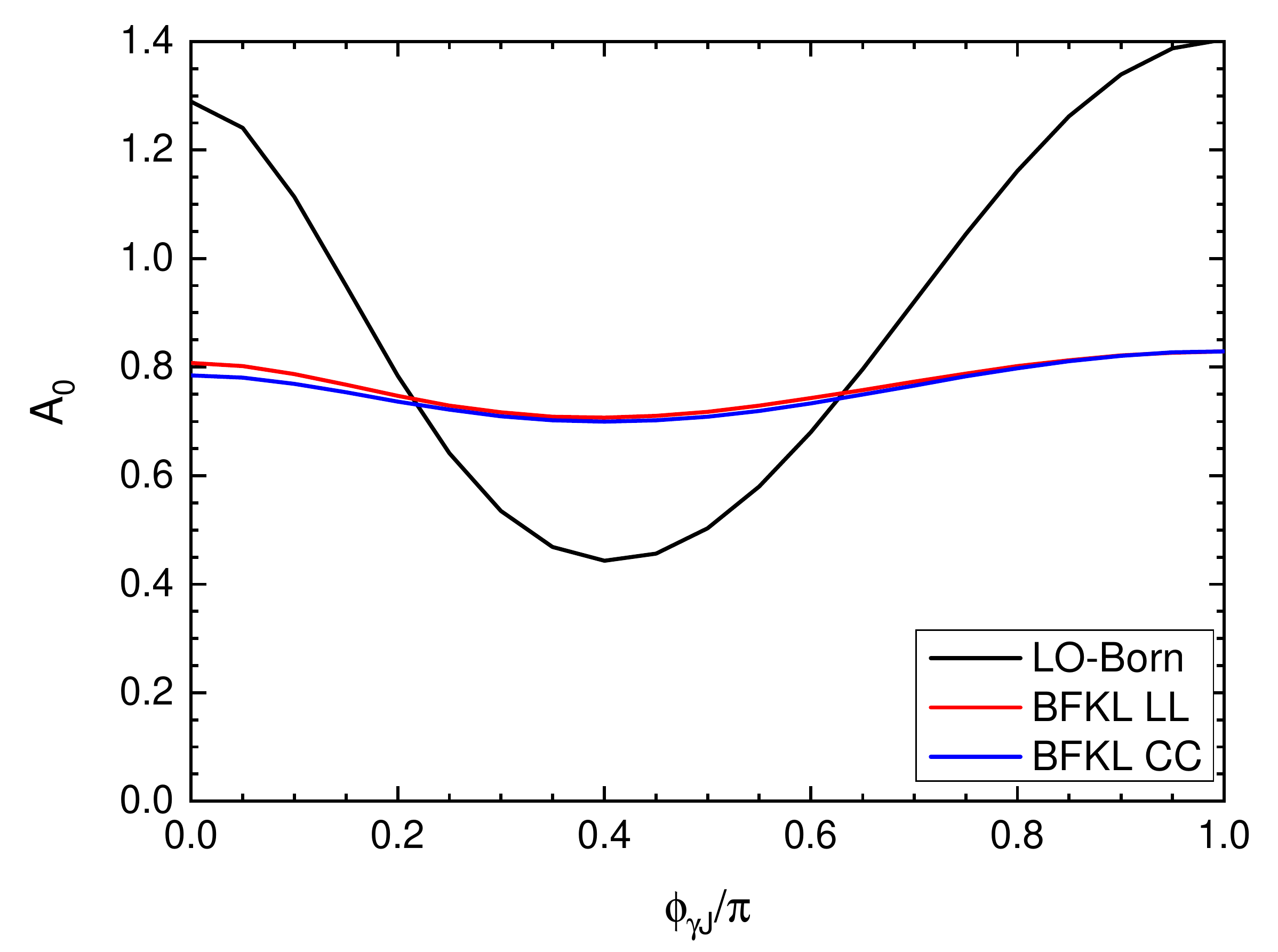}
\includegraphics[width=.46\textwidth]{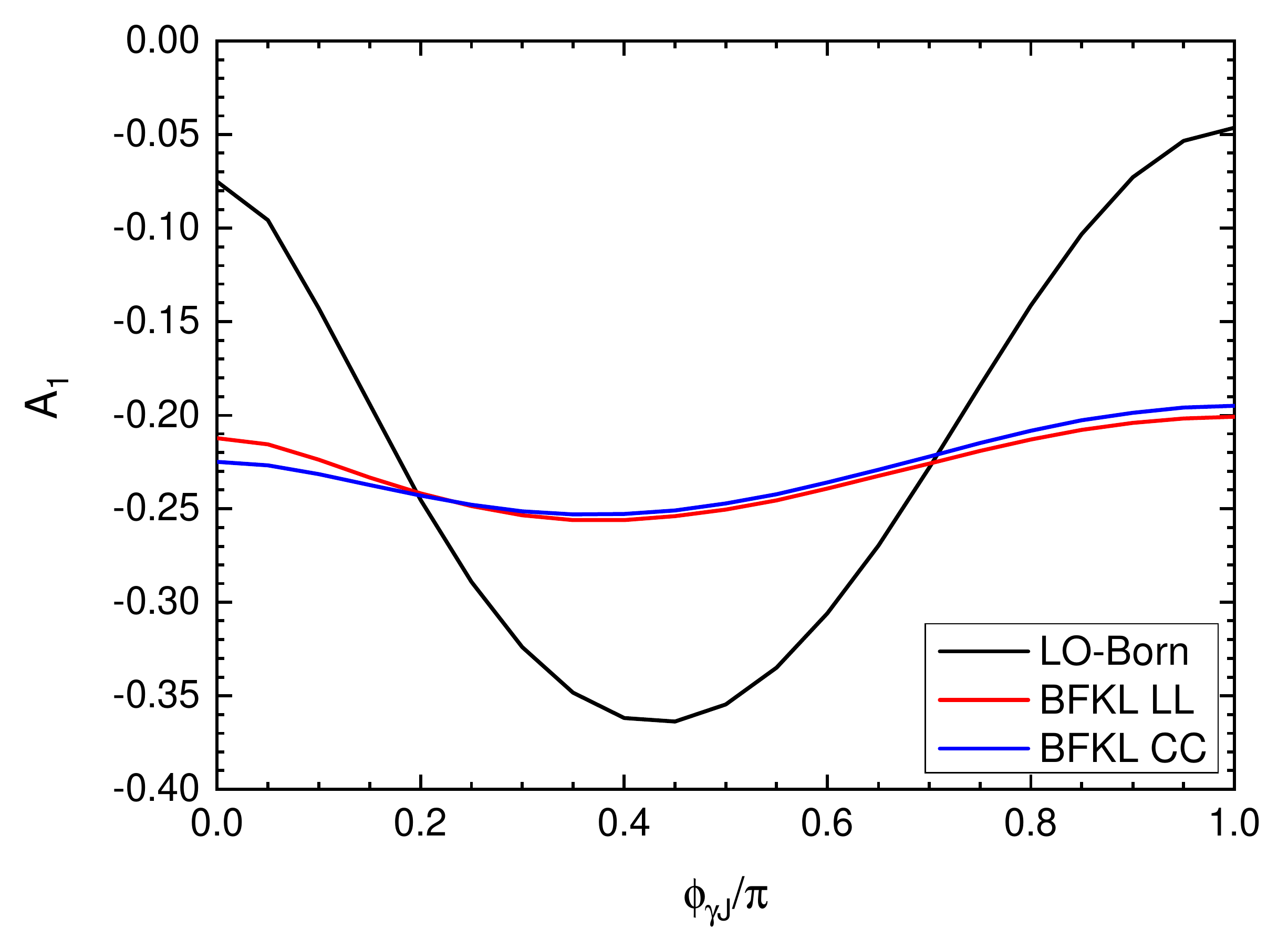}
\includegraphics[width=.46\textwidth]{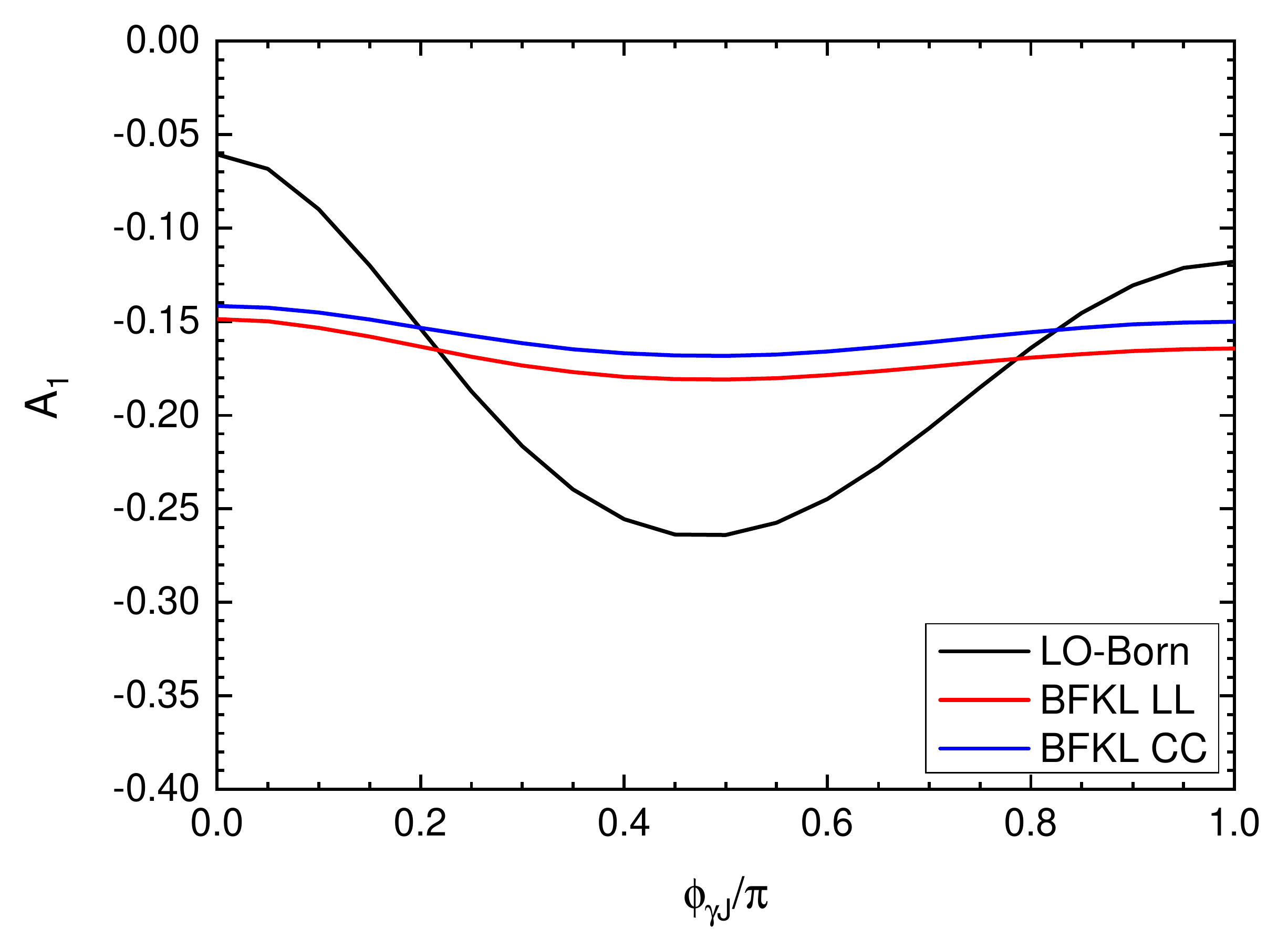}
\includegraphics[width=.46\textwidth]{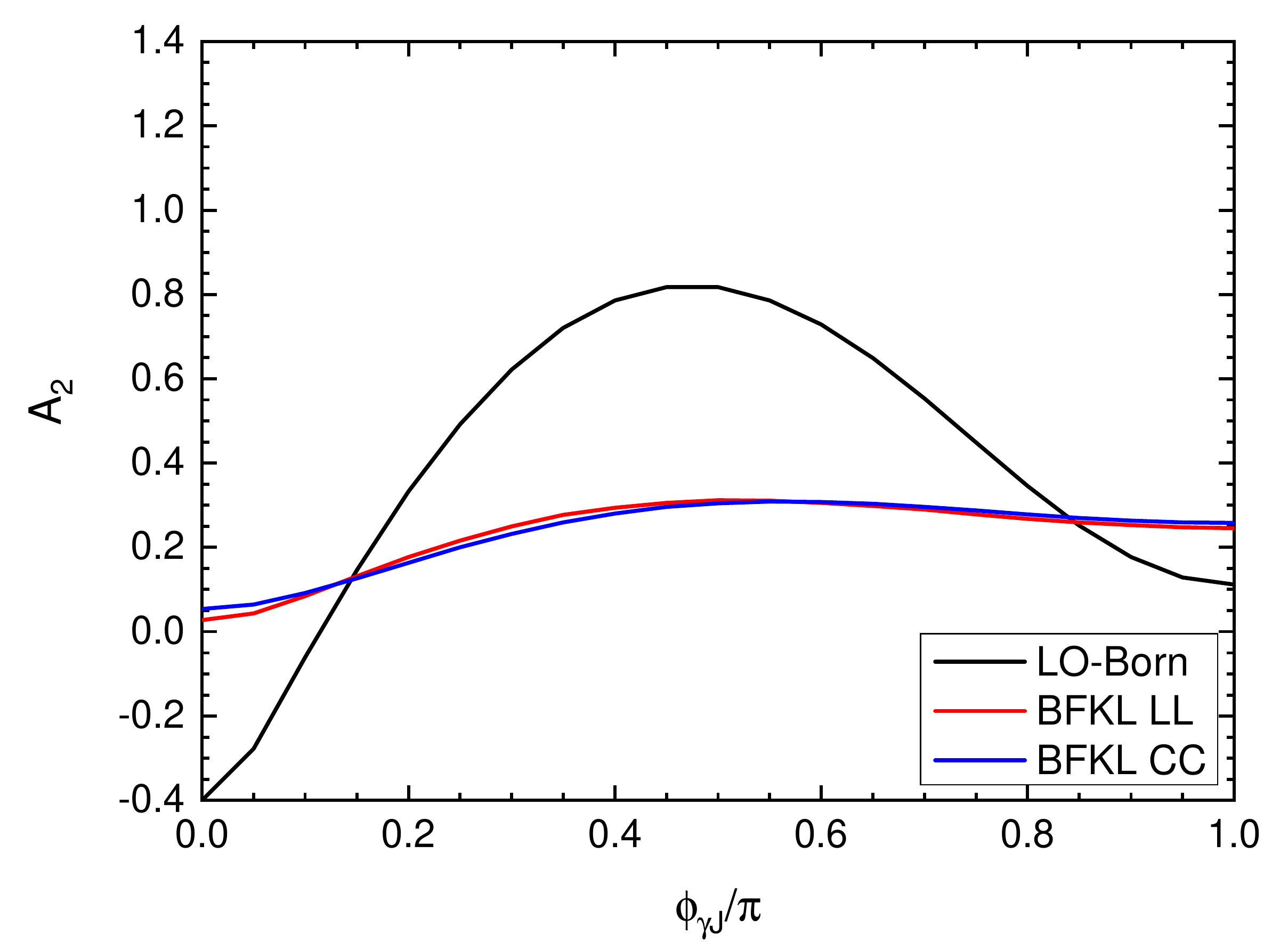}
\includegraphics[width=.46\textwidth]{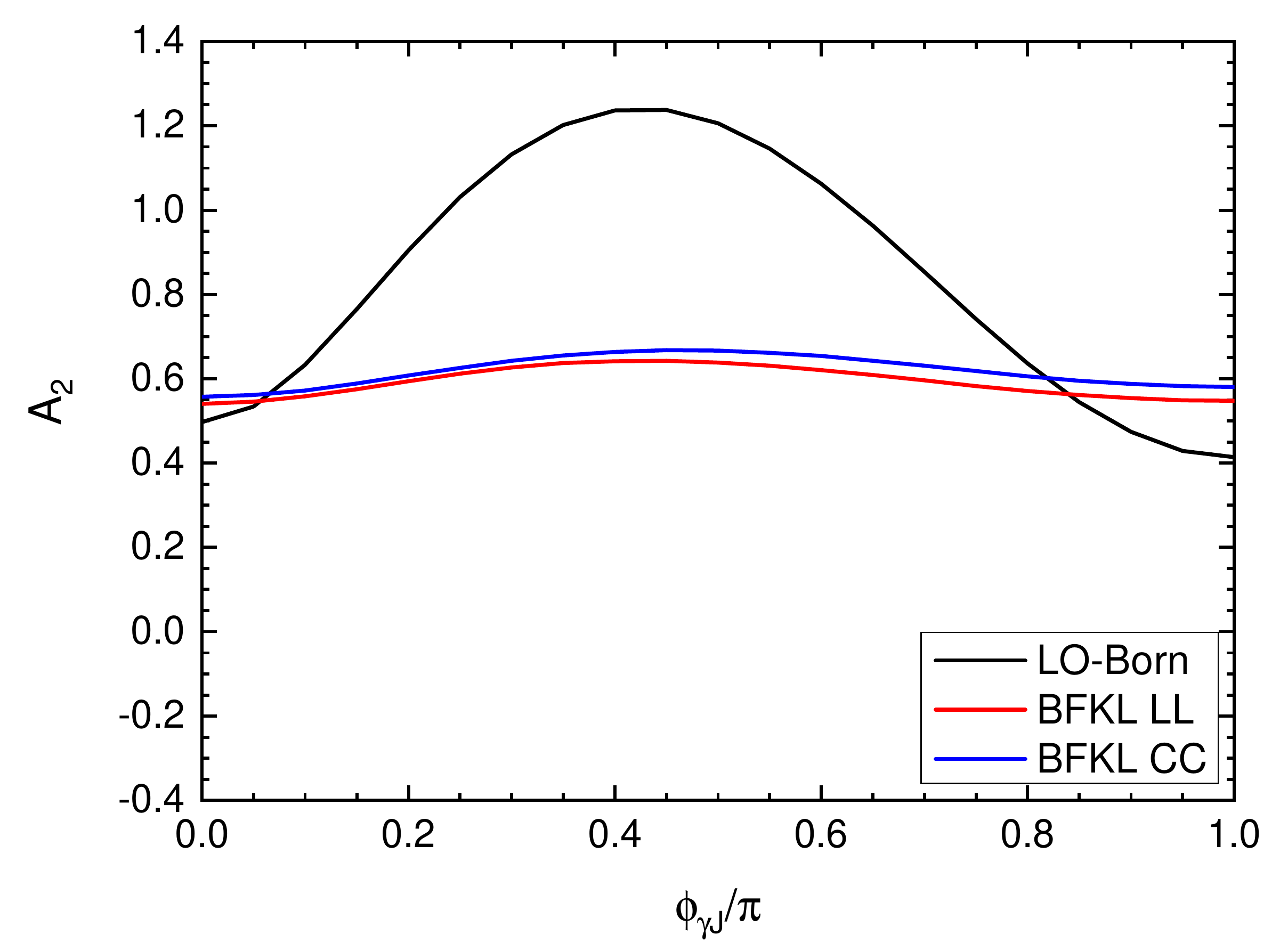}
\includegraphics[width=.46\textwidth]{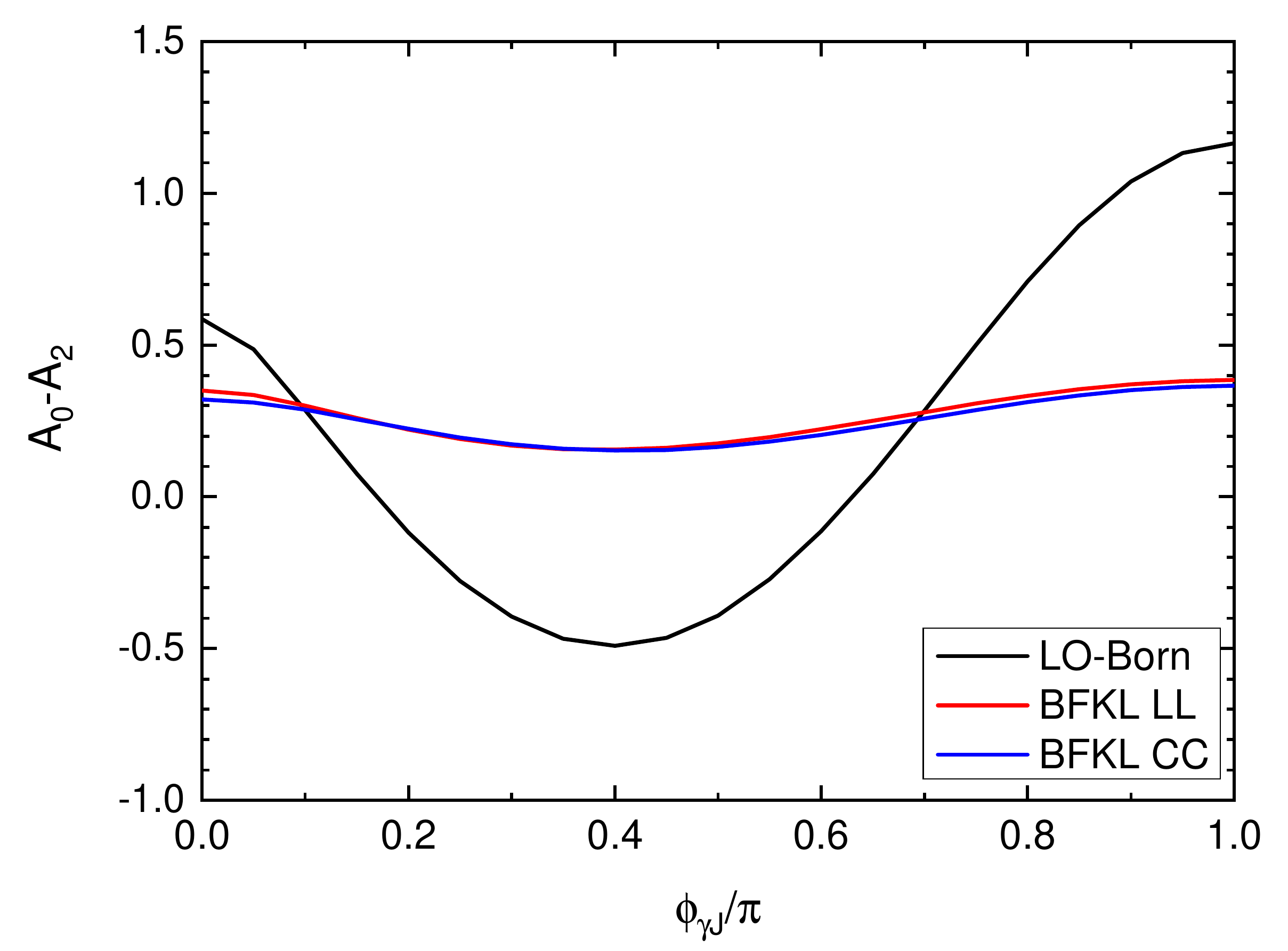}
\includegraphics[width=.46\textwidth]{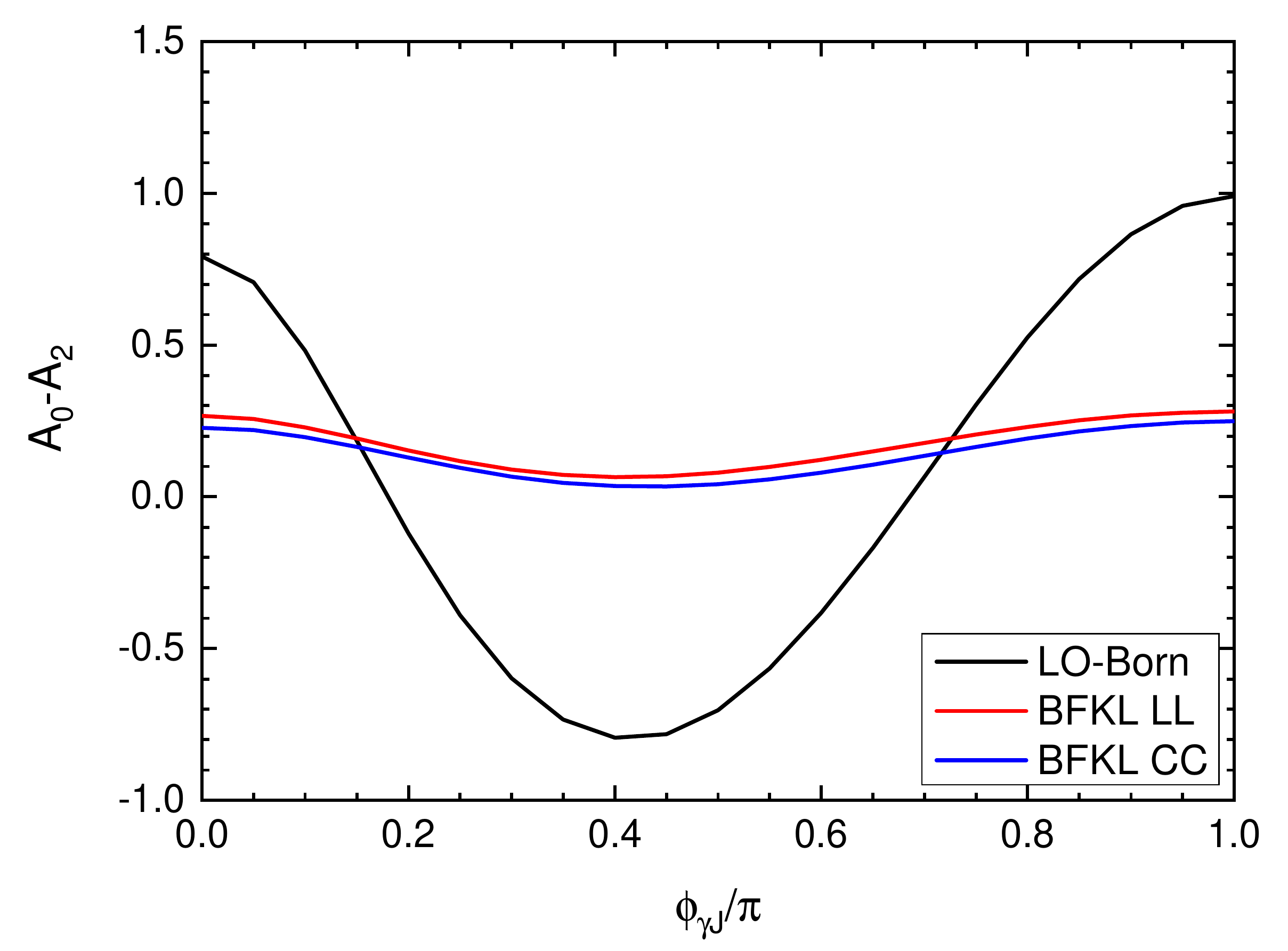}
\end{center}
\caption{The angular coefficients $A_0,A_1$ and $A_2$ as functions of the photon-jet angle $\phi_{\gamma J}$ for the three indicated approximations together with the
Lam-Tung difference $A_0-A_2$. The photon transverse momentum $q_\perp=25~{\rm GeV}$ (left column) and $q_\perp=60~{\rm GeV}$ (right column) while
the other parameters: $p_{J\perp}=30~{\rm GeV}$, $\Delta Y_{\gamma J}=7$ and $M=35~{\rm GeV}$.
}
\label{A_coeff_funct_fi}
\end{figure}

Up to now we have considered only helicity-inclusive quantities which are obtained by averaging over the leptons' distribution. One of the biggest advantage of the DY+jet process, comparing to the MN jet production, is the possibility to investigate the DY lepton angular coefficients $A_i$, defined by eq.~(\ref{A_coeff_def}). In this section we present our analysis of these quantities calculated using the Collins--Soper frame. 

In Fig.~\ref{A_coeff_funct_fi} we show the coefficients $A_0,A_1$ and $A_2$ together with the Lam-Tung difference $A_0-A_2$.
These coefficients are shown as functions of the $\gamma$--jet angle $\phi_{\gamma J}$.
We see a dramatic difference between the LO-Born result which very strongly depends on angle and the BFKL approximations which are almost independent on it. One can conclude that for leptons' angular coefficients the decorrelation coming from the BFKL emissions is almost complete. As before, the LO-Born predictions for the azimuthal dependence are very close to those obtained using the BFKL predictions.

\begin{figure}
\begin{center}
\includegraphics[width=.46\textwidth]{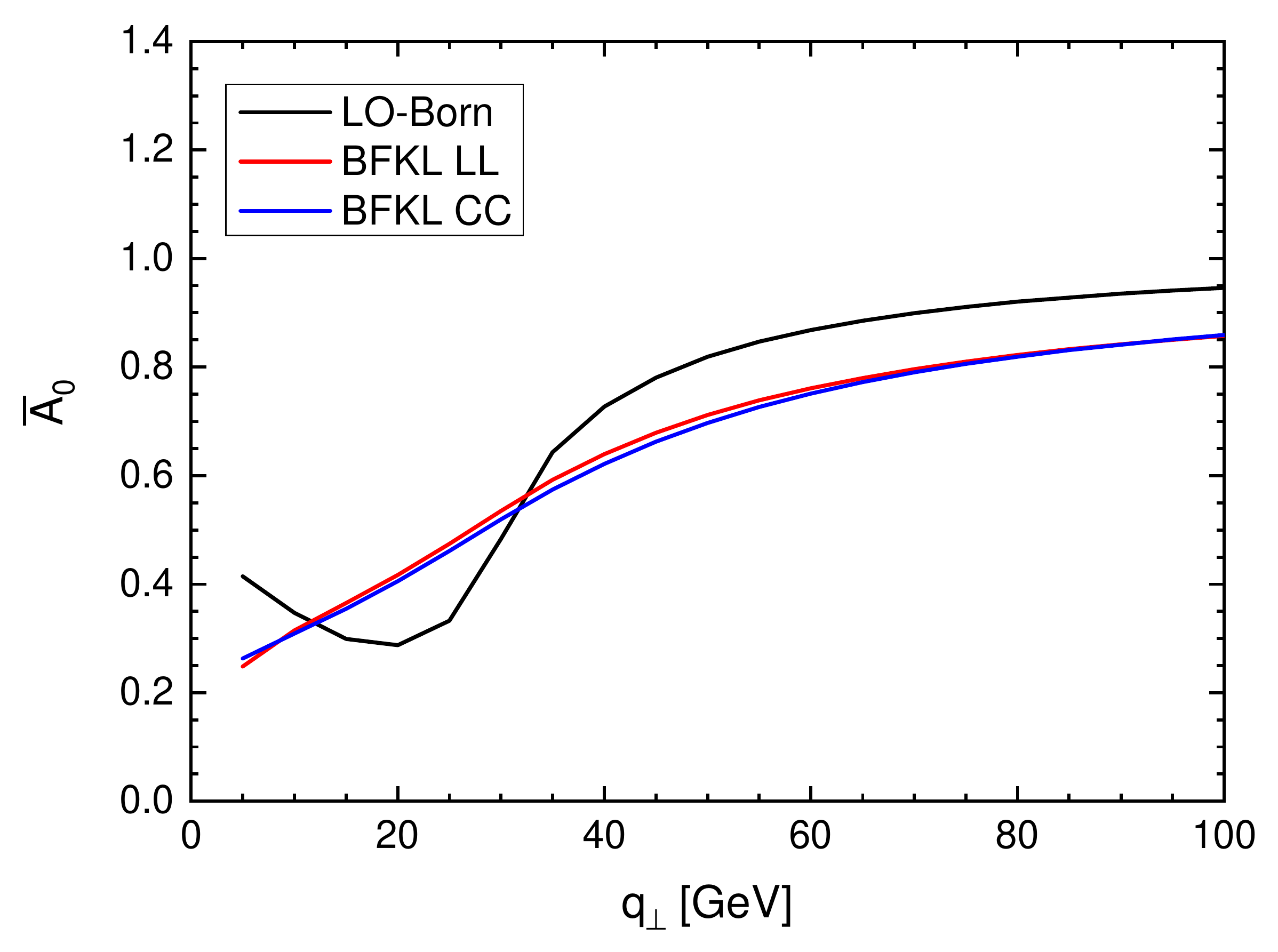}
\includegraphics[width=.46\textwidth]{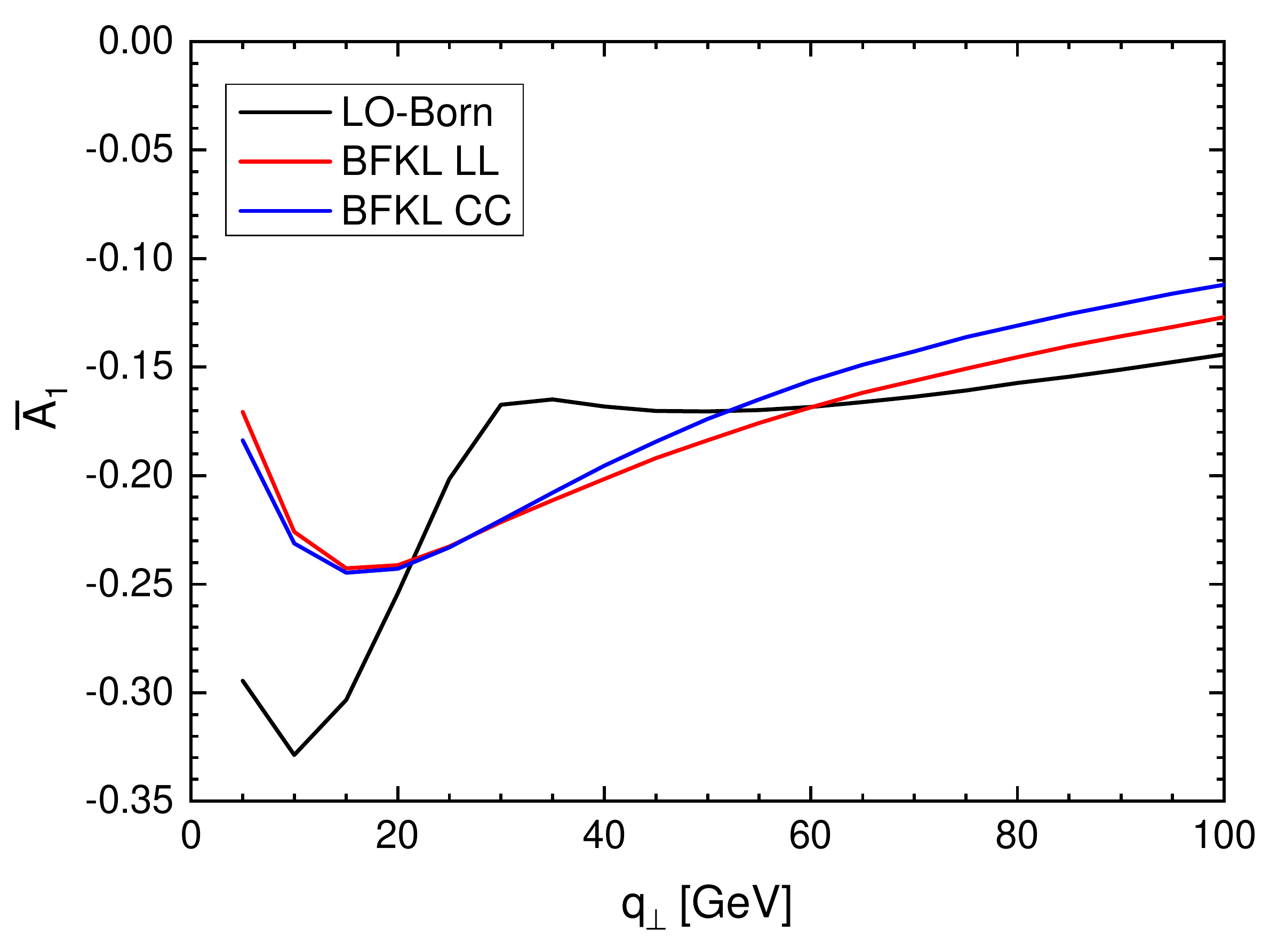}
\includegraphics[width=.46\textwidth]{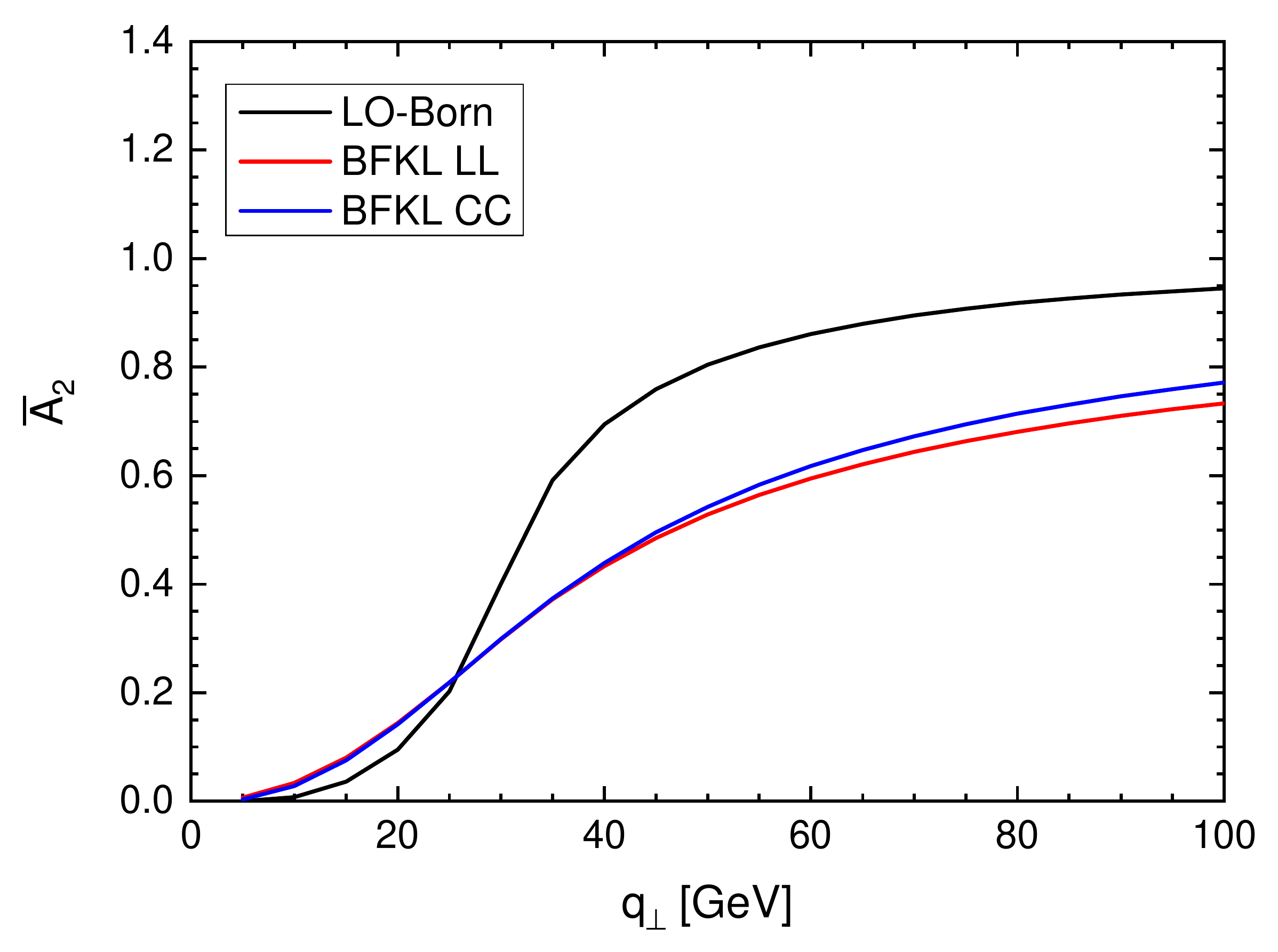}
\includegraphics[width=.46\textwidth]{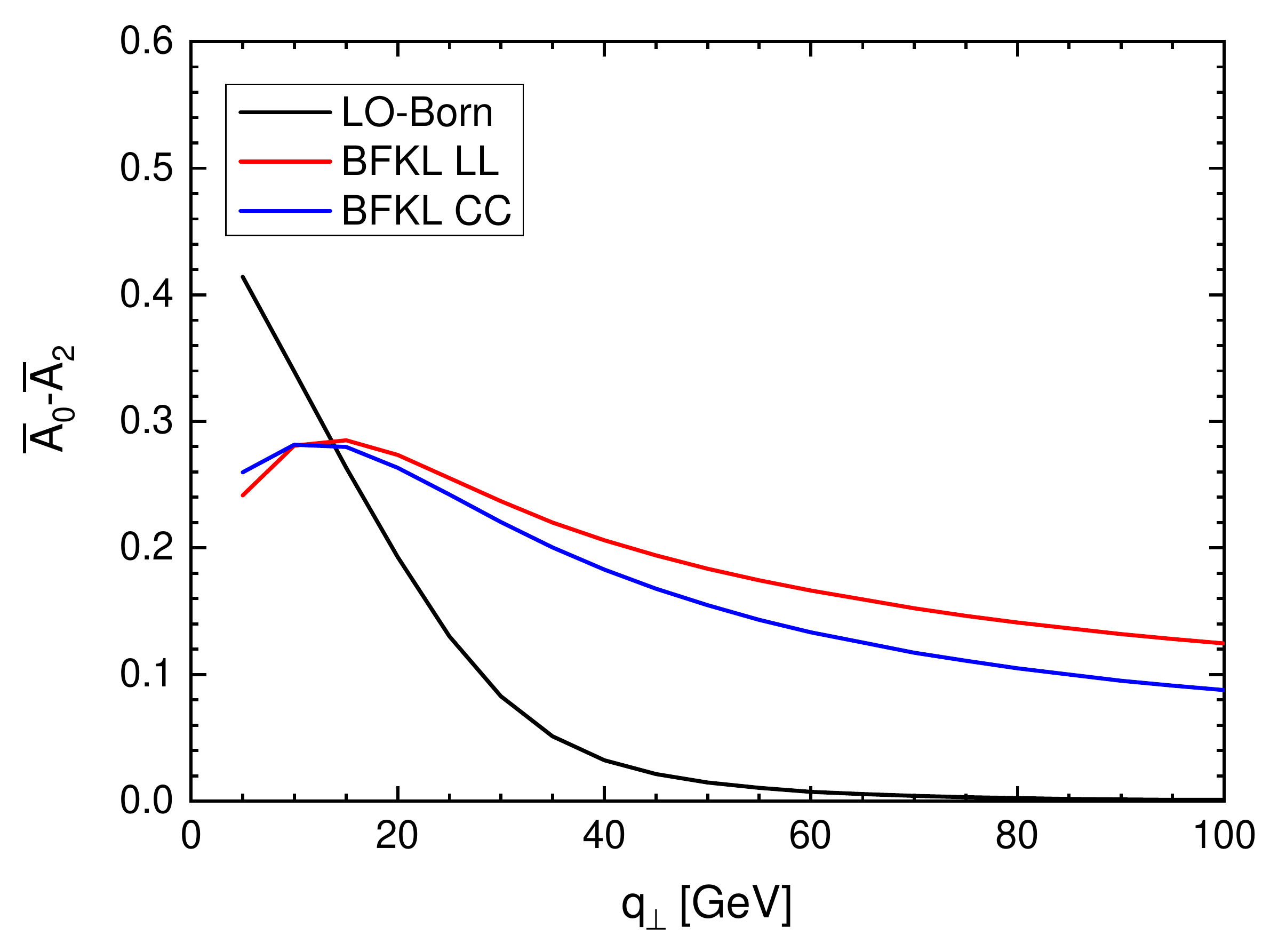}
\end{center}
\caption{The averaged over $\phi_{\gamma J}$ coefficients $\bar{A}_0,\bar{A}_1$ and $\bar{A}_2$ as functions of the photon transverse momentum $q_\perp$
for the three indicated models together with the Lam-Tung difference $\bar{A}_0-\bar{A}_2$. The following parameters are used: $p_{J\perp}=30~{\rm GeV}$, $\Delta Y_{\gamma J}=7$ and $M=35~{\rm GeV}$.
}
\label{A_coeff_funct_qT}
\end{figure}

In order to study the $q_\perp$ and $\Delta Y_{\gamma J}$ dependence of the coefficients $A_i$, it is useful to consider the quantities averaged over the angle $\phi_{\gamma J}$. Therefore, we define the averaged cross sections
\beq\label{eq:5.6ab}
\frac{d\bar\sigma^{(\lambda)}}{dM d\Delta Y_{\gamma J} dq_\perp \, d p_{J \perp}}
=\int_0^{2\pi} d\phi_{\gamma J}\, \frac{d\sigma^{(\lambda)}}{dM d\Delta Y_{\gamma J} dq_\perp \, d p_{J \perp}d\phi_{\gamma J}}\, .
\eeq
Then the $\bar A_i$'s defined by eqs.~(\ref{A_coeff_def}) are computed using the averaged $d\bar\sigma^{(\lambda)}$'s. The calculation of (\ref{eq:5.6ab}) for 
the BFKL cross section (\ref{fourier_exp_of_master_form}) is particularly simple since all the Fourier coefficients with $m\ge 1$ vanish and 
\beq
\frac{d\bar\sigma^{(\lambda)}}{dM d\Delta Y_{\gamma J} dq_\perp \, d p_{J \perp}}\, = 2\pi \, \mathcal{I}^{(\lambda)}_0(M, \Delta Y_{\gamma J}, q_\perp, p_{J \perp}).
\eeq
In Fig.~\ref{A_coeff_funct_qT} we show the averaged coefficients $\bar A_i$'s as functions of $q_\perp$. The Lam--Tung observable is particularly interesting. 
In the LO-Born approximation it decreases rapidly with $q_\perp$, so that it vanishes when $q_\perp$ is substantially larger than $ p_{J\perp}$. It is easy to understand since violation of the Lam--Tung relation is caused in this process by the transverse momentum transfer from the forward jet to the DY impact factor. When $ p_{J\perp}$ is substantially smaller than $q_\perp$, this momentum transfer is negligible and the Lam-Tung relation is satisfied. On the other hand, the BFKL emissions provide large transverse momentum transfer to the DY impact factor even when $p_{J\perp}$ is small comparing to $q_\perp$.

\begin{figure}
\begin{center}
\includegraphics[width=.46\textwidth]{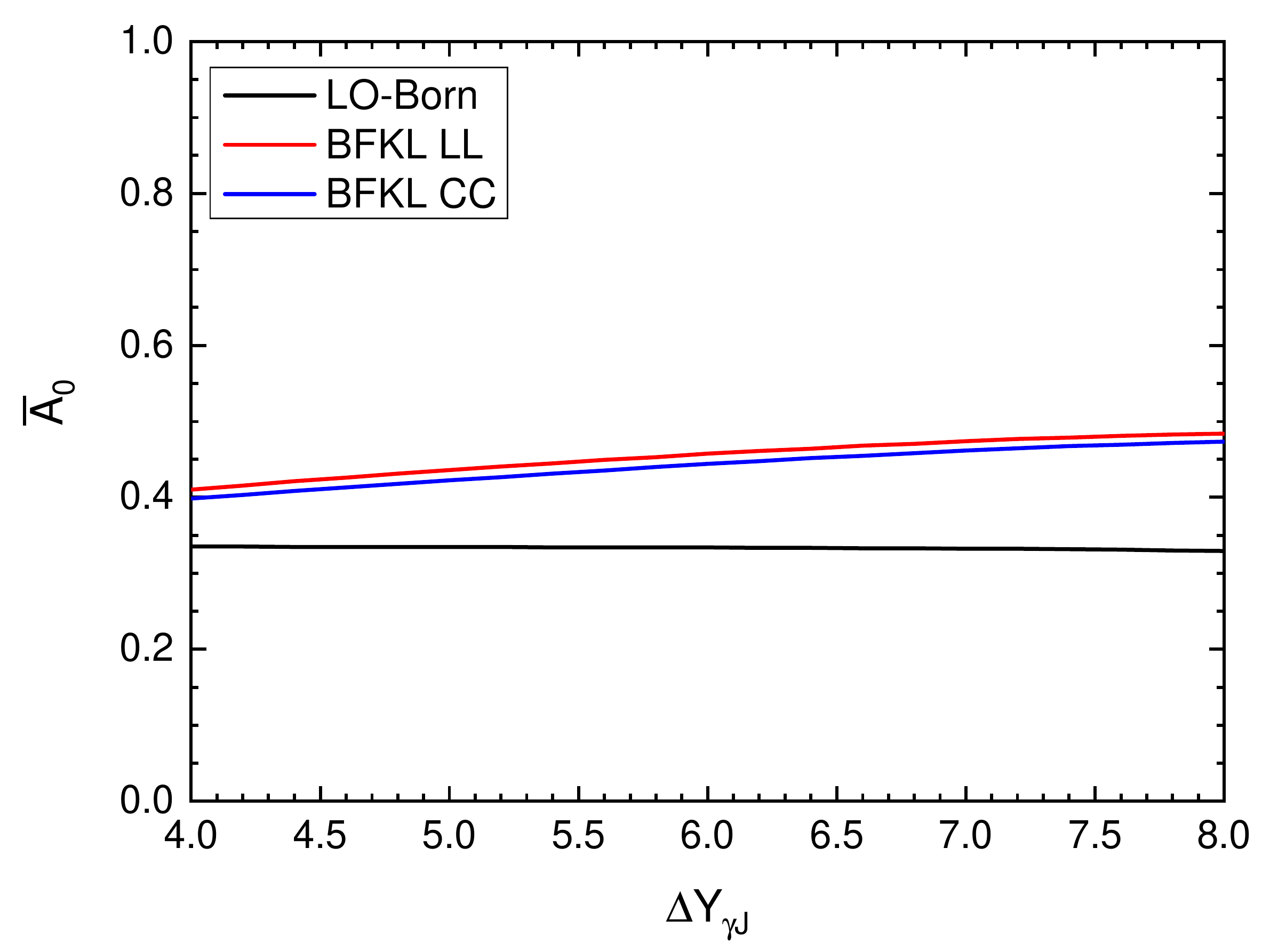}
\includegraphics[width=.46\textwidth]{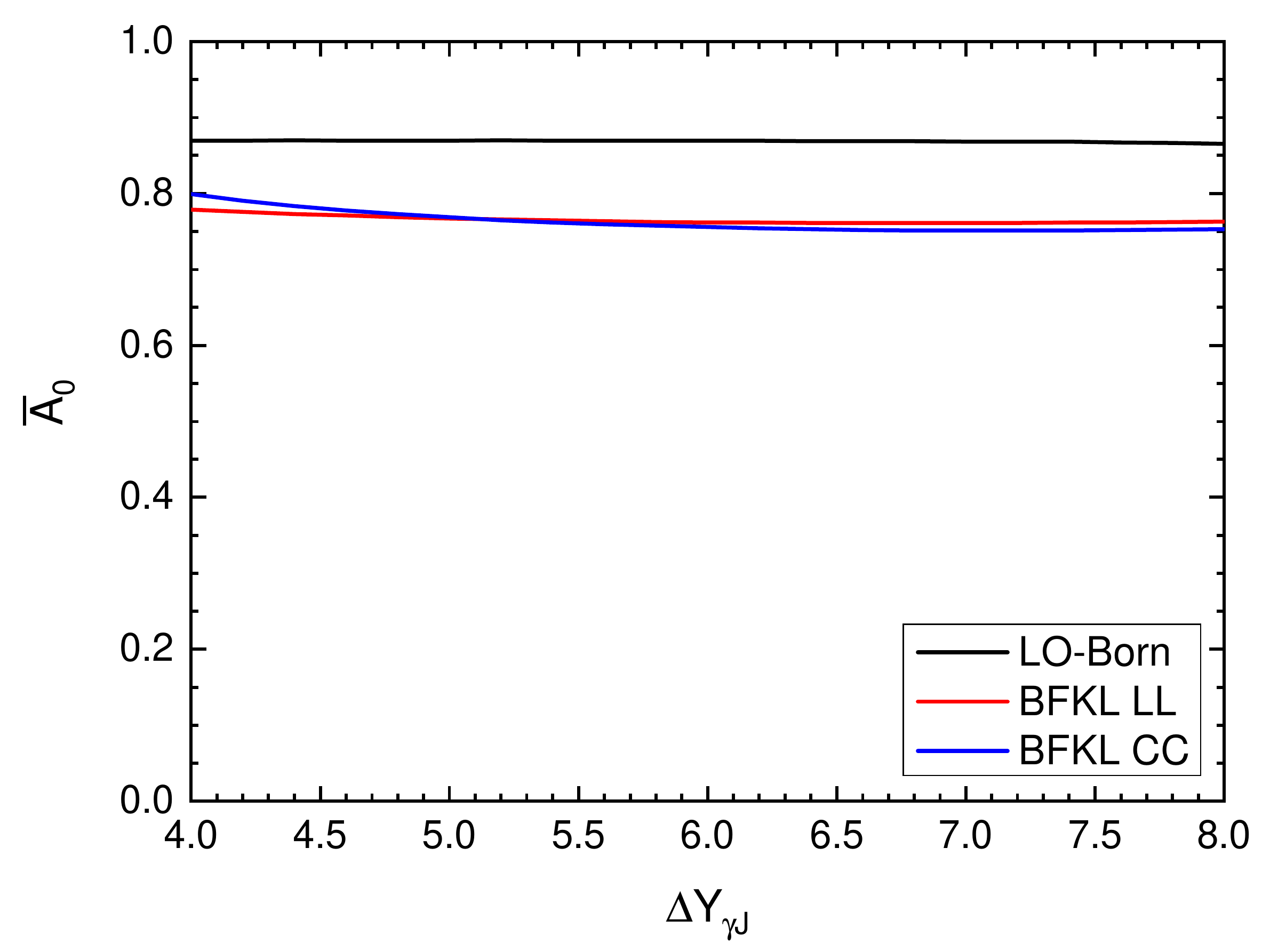}
\includegraphics[width=.46\textwidth]{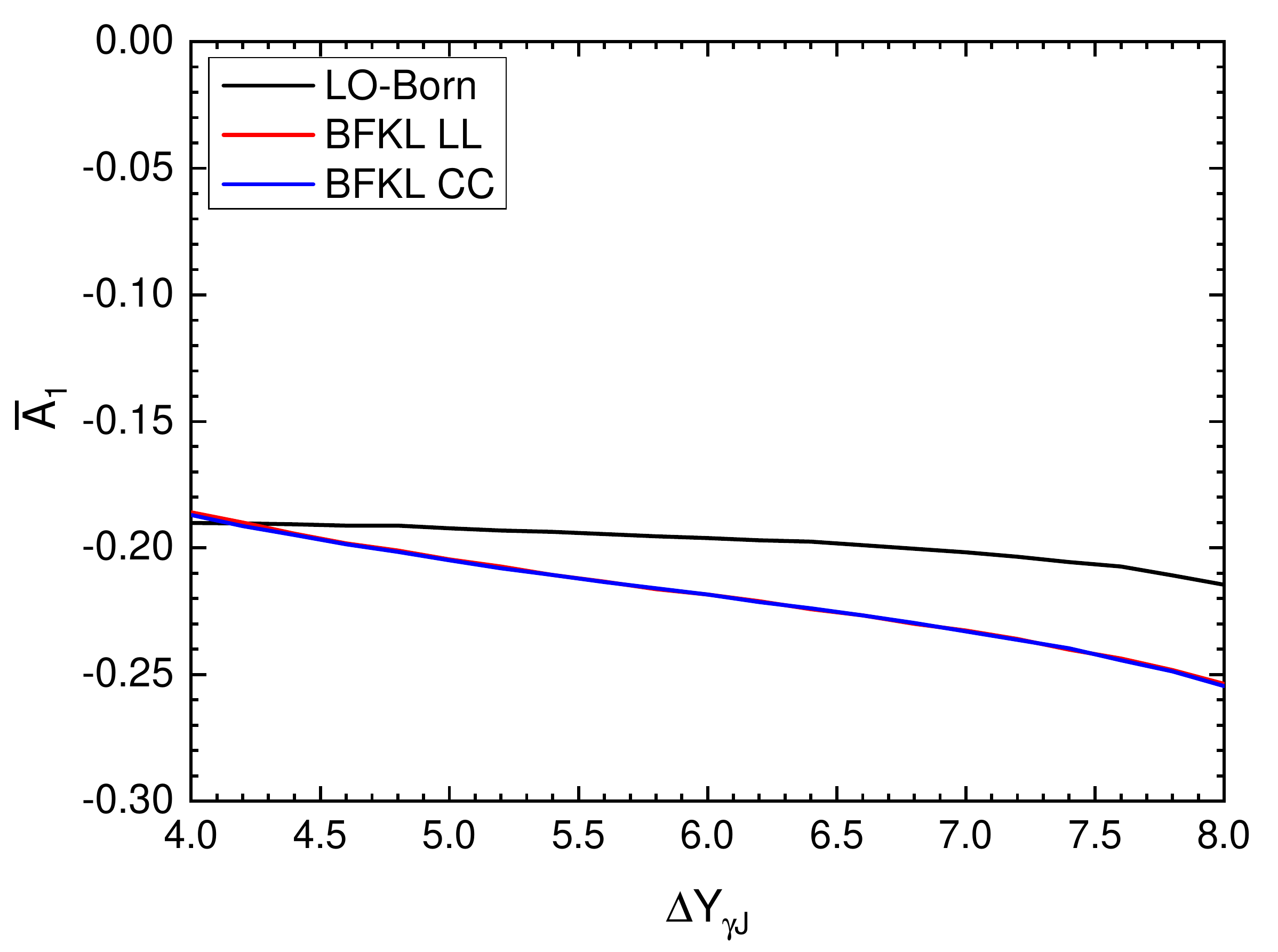}
\includegraphics[width=.46\textwidth]{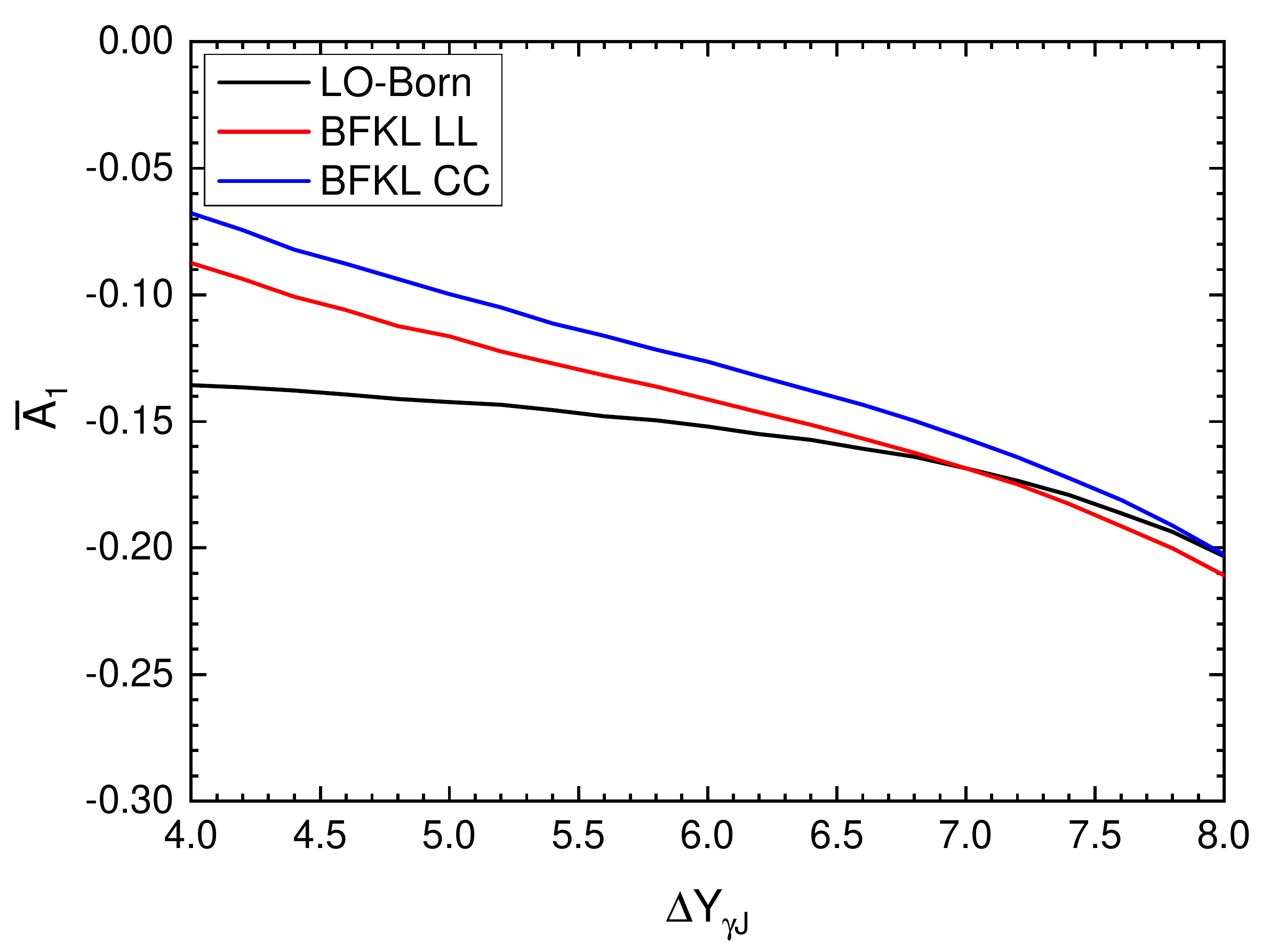}
\includegraphics[width=.46\textwidth]{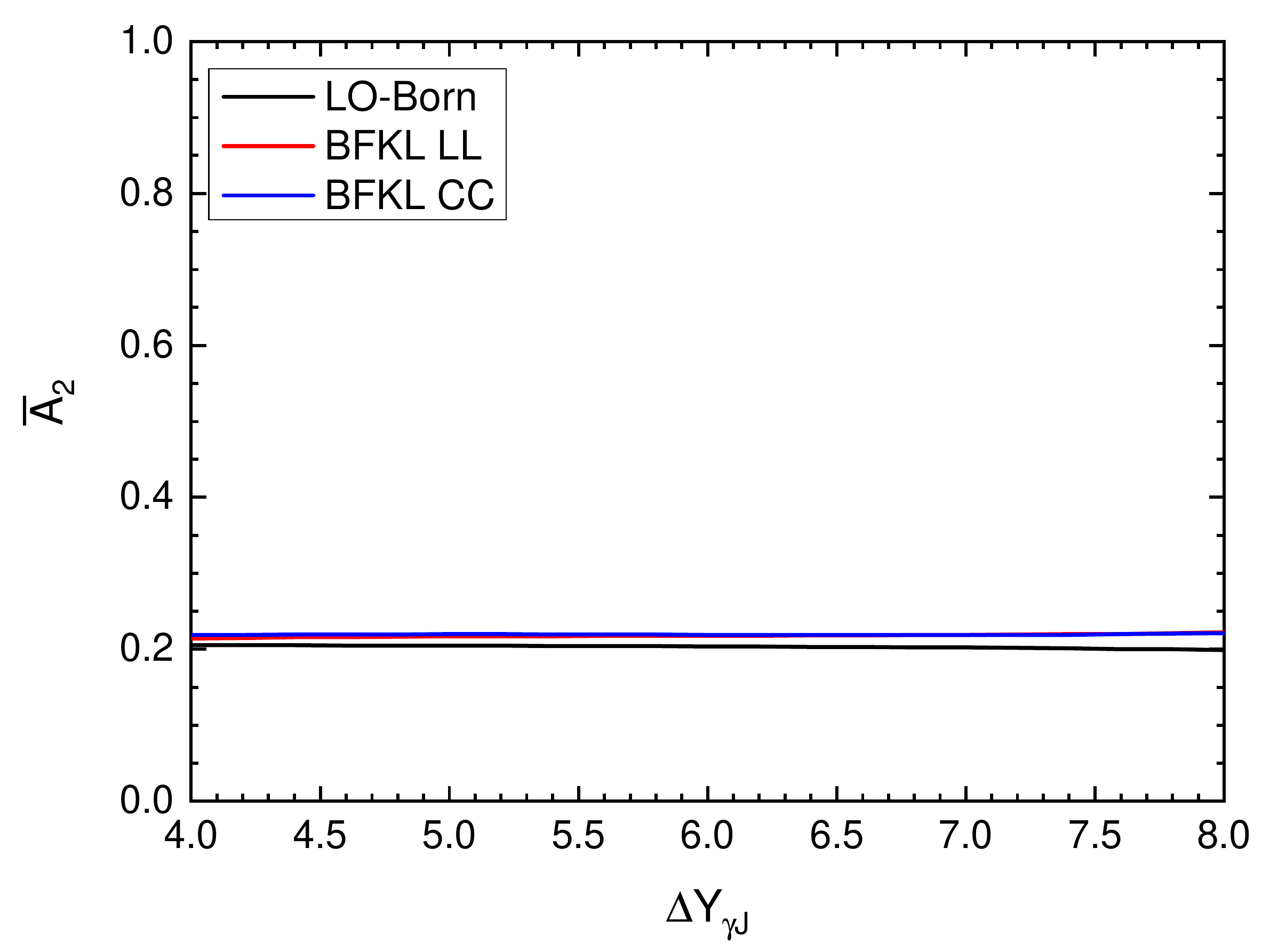}
\includegraphics[width=.46\textwidth]{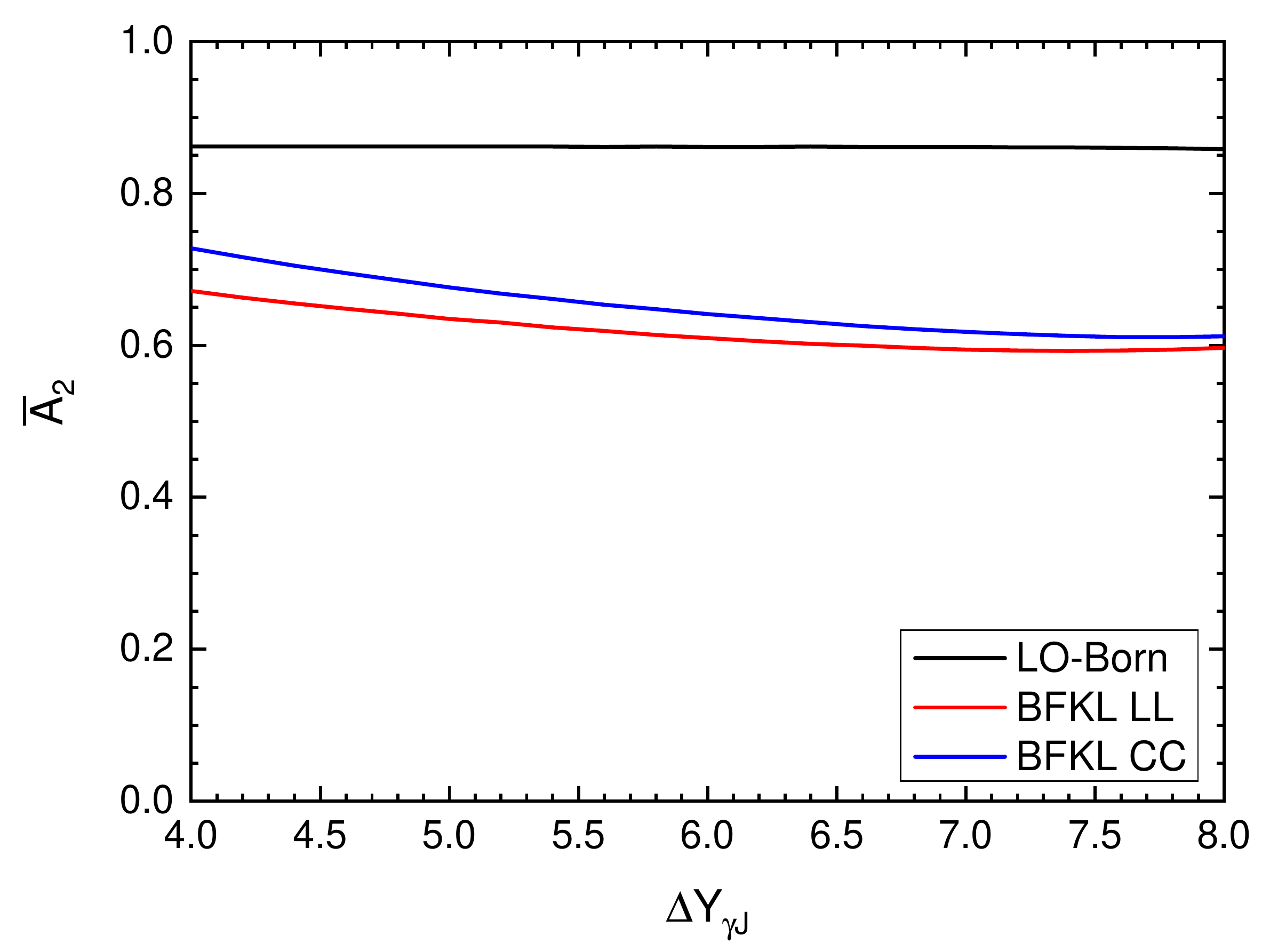}
\includegraphics[width=.46\textwidth]{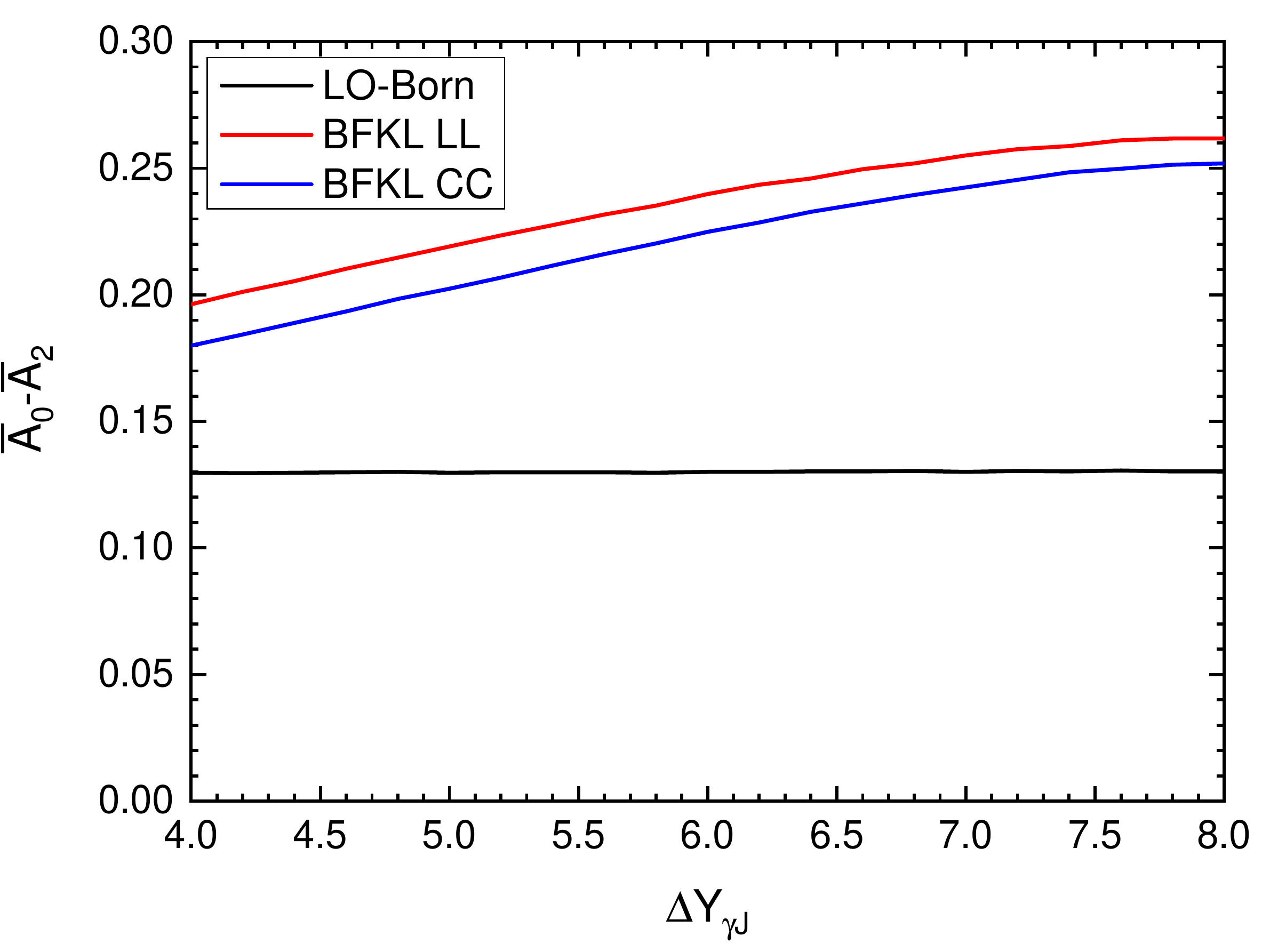}
\includegraphics[width=.46\textwidth]{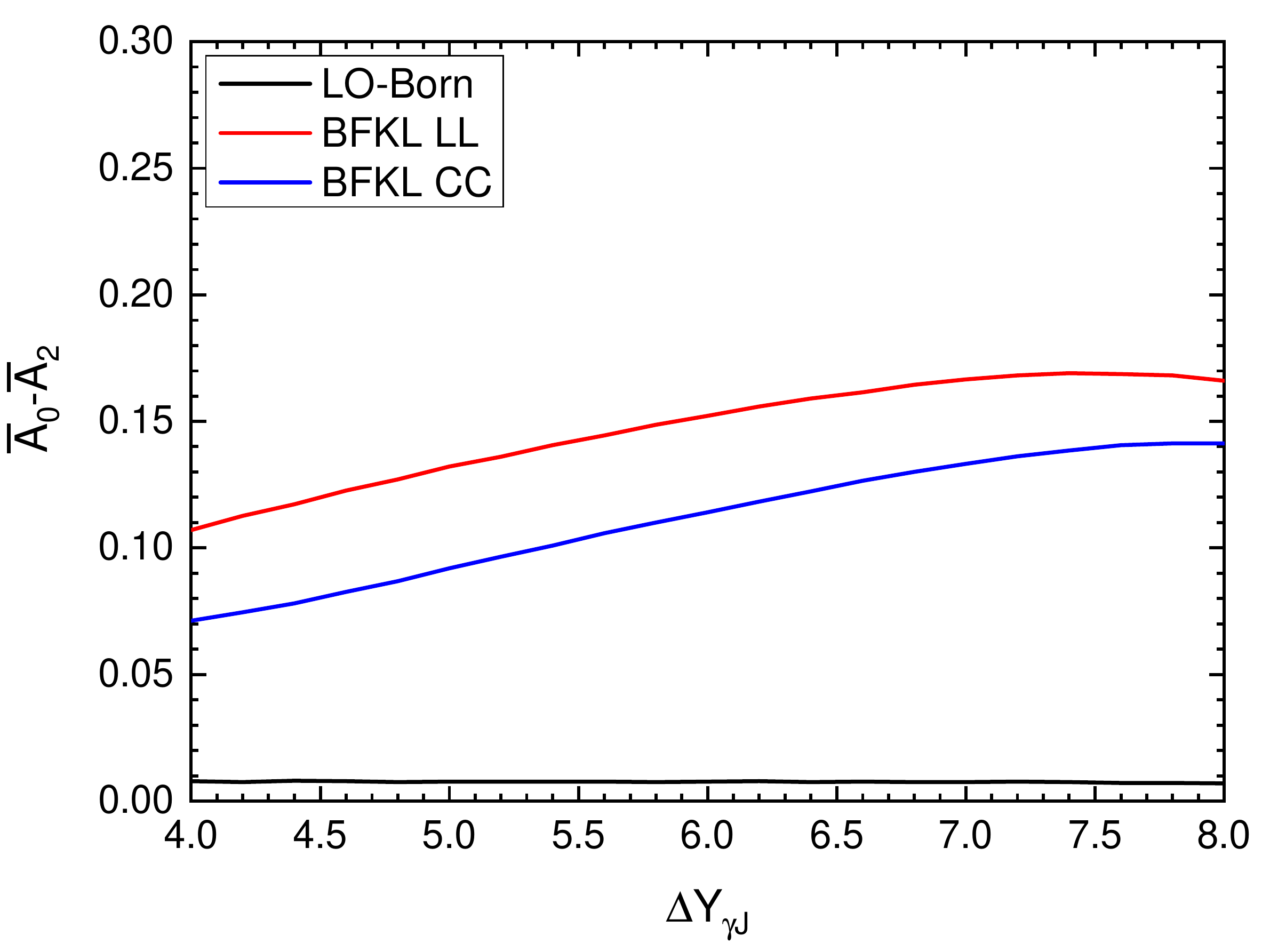}
\end{center}
\caption{The averaged over $\phi_{\gamma J}$ coefficients $\bar{A}_0,\bar{A}_1$ and $\bar{A}_2$ as functions of the photon--jet rapidity difference $\Delta Y_{\gamma J}$
for the three indicated models together with the Lam-Tung difference $\bar{A}_0-\bar{A}_2$. The photon transverse momentum $q_\perp=25~{\rm GeV}$ (left column) and $q_\perp=60~{\rm GeV}$ (right column) while the other parameters: $p_{J\perp}=30~{\rm GeV}$, $\Delta Y_{\gamma J}=7$ and $M=35~{\rm GeV}$.
}
\label{A_coeff_funct_Y}
\end{figure}

In Fig.~\ref{A_coeff_funct_Y}, the mean coefficients $\bar A_i$ are shown as a function of $\Delta Y_{\gamma J}$ for $q_\perp=25$ GeV (left column) and $q_\perp=60$ GeV (right column). We see again a significant difference between the LO-Born and the BFKL approximations.

\section{Summary and outlook}

W proposed a new process to study the BFKL dynamics in high energy hadronic collisions -- the Drell-Yan (DY) plus jet production. In this process, the DY photon with large rapidity difference with respect to the backward jet should be tagged. The process is inclusive in a sense that the rapidity space between the forward photon and the backward jet can be populated by minijets which are described as the BFKL radiation. As in the classical Mueller-Navelet process with two jets separated by a large rapidity interval, we propose to look at decorrelation of the azimuthal angle between the DY boson and the forward jet. For the estimation of the size of this effect, we use the formalism with the BFKL kernel in two approximations; the leading logarithmic (LL) and the approximation with consistency conditions (CC) which takes into account majority of the next-to-leading logarithmic corrections to the BFKL radiation. The jet and photon impact factors were taken in the lowest order approximation.

The presented numerical results show a significant angular decorrelation with respect to the Born approximation for the BFKL kernel, which is observed for all considered values of photon transverse momentum. The found decorrelation is stronger than for the Mueller-Navelet jets due to more complicated final state with one more particle, being the tagged DY boson. We also presented numerical results on the angular coefficients of the DY lepton pair which provide an additional experimental opportunity to test the effect of the BFKL dynamics in the proposed process. In particular, these coefficients allow to study the Lam-Tung relation (\ref{Lam-Tung_combination}) which is strongly sensitive to the transverse momentum transfer to the DY impact factor. For this reason, the study of the angular coefficients of the DY pair in the BFKL framework is highly interesting.

As an outlook, it would be very interesting to analyse the DY$\,+\,$jet production in full NLO and NLL setting for the photon/jet impact factors and the BFKL kernel. We hope to return to this problem in future.

\section*{Acknowledgments}
TS acknowledges the Mobility Plus grant of the Ministry of Science and Higher Education of Poland and thanks the Brookhaven National Laboratory for hospitality and support. This work was also supported by the National Science Center, Poland, Grants No. 2015/17/B/ST2/01838, DEC-2014/13/B/ST2/02486 and 2017/27/B/ST2/02755.

\appendix

\bibliographystyle{JHEP}
\bibliography{mybib}

\end{document}